\documentclass[a4paper,11pt]{article}
\usepackage{a4wide,amsmath,amssymb,bbm}
\usepackage[english]{babel}

\usepackage{color}
\usepackage{slashed}
\usepackage[normalem]{ulem}
\usepackage{cite}
\usepackage[bottom]{footmisc}

\usepackage[pdfencoding=auto,psdextra]{hyperref}
\usepackage{enumerate}
\hypersetup{
    colorlinks,
    linkcolor={black},
    citecolor={black},
    urlcolor={black}
}

\makeatletter
 \let\old@startsection=\@startsection
 \let\oldl@section=\l@section
 \renewcommand{\@startsection}[6]{\old@startsection{#1}{#2}{#3}{#4}{#5}{#6\mathversion{bold}}}
 \renewcommand{\l@section}[2]{\oldl@section{\mathversion{bold}#1}{#2}}
\makeatother

\makeatletter
\DeclareFontFamily{OMX}{MnSymbolE}{}
\DeclareSymbolFont{MnLargeSymbols}{OMX}{MnSymbolE}{m}{n}
\SetSymbolFont{MnLargeSymbols}{bold}{OMX}{MnSymbolE}{b}{n}
\DeclareFontShape{OMX}{MnSymbolE}{m}{n}{
    <-6>  MnSymbolE5
   <6-7>  MnSymbolE6
   <7-8>  MnSymbolE7
   <8-9>  MnSymbolE8
   <9-10> MnSymbolE9
  <10-12> MnSymbolE10
  <12->   MnSymbolE12
}{}
\DeclareFontShape{OMX}{MnSymbolE}{b}{n}{
    <-6>  MnSymbolE-Bold5
   <6-7>  MnSymbolE-Bold6
   <7-8>  MnSymbolE-Bold7
   <8-9>  MnSymbolE-Bold8
   <9-10> MnSymbolE-Bold9
  <10-12> MnSymbolE-Bold10
  <12->   MnSymbolE-Bold12
}{}

\let\llangle\@undefined
\let\rrangle\@undefined
\DeclareMathDelimiter{\llangle}{\mathopen}%
                     {MnLargeSymbols}{'164}{MnLargeSymbols}{'164}
\DeclareMathDelimiter{\rrangle}{\mathclose}%
                     {MnLargeSymbols}{'171}{MnLargeSymbols}{'171}
\makeatother

\usepackage{mathrsfs}

\usepackage{comment}
\usepackage{braket}

\newcommand{\Tr}{\mathrm{Tr}}
\newcommand{\be}{\begin{equation}}
\newcommand{\ee}{\end{equation}}
\newcommand{\alg}{\mathfrak}

\newcommand{\Ad}{\operatorname{ad}}
\newcommand{\AD}{\operatorname{Ad}}

\makeatletter

\@addtoreset{equation}{section} \makeatother

\begin{document}

\phantom{.}

\vspace{30pt}

\begin{center}
{\huge{\bf  Homogeneous Yang-Baxter deformations\\[5pt] 
as   undeformed yet twisted  models }}

\vspace{80pt}

Riccardo Borsato, \ \ Sibylle Driezen \ \ and \ \ J. Luis Miramontes

\vspace{15pt}

{
\small {{\it 
Instituto Gallego de F\'isica de Altas Energ\'ias (IGFAE),\\[2pt]
and\\[2pt]
Departamento de F\'\i sica de Part\'\i culas,\\[7pt]
Universidad de  Santiago de Compostela
\\[3pt]
15782 Santiago de Compostela, Spain}\\[5pt]
\vspace{12pt}
\texttt{riccardo.borsato@usc.es, sib.driezen@gmail.com, jluis.miramontes@usc.es}}}\\

\vspace{100pt}

{\bf Abstract}
\end{center}
\noindent
The homogeneous Yang-Baxter deformation is part of a larger web of integrable deformations and dualities that recently have been studied with motivations in integrable $\sigma$-models, solution-generating techniques in supergravity and Double Field Theory, and possible generalisations of the AdS/CFT correspondence. The $\sigma$-models obtained by the homogeneous Yang-Baxter deformation with periodic boundary conditions on the worldsheet  are on-shell equivalent to undeformed models, yet with twisted boundary conditions. While this has been known for some time, the expression provided so far for the twist features non-localities (in terms of the degrees of freedom of the deformed model) that prevent practical calculations, and in particular the construction of the classical spectral curve. We solve this problem by rewriting the equation defining the twist in terms of the degrees of freedom of the undeformed yet twisted model, and we show that we are able to solve it in full generality. Remarkably, this solution is a local expression. We discuss the consequences of the twist at the level of the monodromy matrix and of the classical spectral curve, analysing in particular the concrete examples of abelian, almost abelian and Jordanian deformations of the Yang-Baxter  class.

\pagebreak 
\tableofcontents

\setcounter{page}{1}
\newcounter{nameOfYourChoice}

\section{Introduction}
In  recent years there has been a lot of activity in the study of integrability-preserving deformations of 2-dimensional $\sigma$-models, see~\cite{Orlando:2019his,Klimcik:2021bjy,Hoare:2021dix} for recent reviews. In this paper we will deal with the so-called ``homogeneous Yang-Baxter'' (hYB) deformation~\cite{Kawaguchi:2014qwa,vanTongeren:2015soa}, which is a variant of the ``inhomogeneous Yang-Baxter'' deformation (sometimes  called $\eta$-deformation) of~\cite{Klimcik:2002zj,Klimcik:2008eq,Delduc:2013fga,Delduc:2013qra}. In principle, the hYB deformation can  be applied to any $\sigma$-model with a group $G$ of isometries \cite{Bakhmatov:2017joy,Bakhmatov:2018apn,Borsato:2018idb}, and it is generated by a linear operator $R$ on $\alg g$ (the Lie algebra of $G$) that solves the classical Yang-Baxter equation. In the case of the inhomogeneous deformation $R$ solves the \emph{modified} classical Yang-Baxter equation instead. Both are integrable deformations in the sense that, when the original $\sigma$-model that one wants to deform admits a Lax connection, it is possible to construct  Lax connections also for the deformed models. When this happens, one may  conclude that the deformed models are classically integrable at least in the weak sense.\footnote{A strong version of integrability requires also the proof that the conserved charges extracted from the monodromy matrix are in involution with each other, see~\cite{Klimcik:2020fhs} for a study in this context.}

The inhomogeneous deformation is known to have an interpretation in terms of a quantum group deformation~\cite{Delduc:2013fga,Delduc:2016ihq}. In this case, the set of choices for $R$ is generically very limited, albeit the fact that more than one possibility is available turned out to be crucial in the study of deformations of supercoset $\sigma$-models~\cite{Hoare:2018ngg}.
In this respect, the hYB deformation gives more freedom, because for a generic $\alg g$ there may be a rich class of solutions for $R$, and each of these solutions will then give rise to a different deformation with its unique features. In particular, the models may differ in which isometries are broken under the deformation and which ones survive.

Both the homogeneous and the inhomogeneous Yang-Baxter deformations are actually related to generalisations of T-duality transformations, repectively to non-abelian T-duality~\cite{delaOssa:1992vci,Gasperini:1993nz,Giveon:1993ai} and Poisson-Lie T-duality~\cite{Klimcik:1995ux,Klimcik:1995dy}. In fact, it is possible to view the hYB deformation as an interpolation between the original model and its non-abelian T-dual~\cite{Hoare:2016wsk,Borsato:2016pas}. Moreover, the inhomogeneous deformation is related via Poisson-Lie duality to the $\lambda$-deformation of~\cite{Sfetsos:2013wia,Hollowood:2014rla,Hollowood:2014qma},  as discussed in~\cite{Vicedo:2015pna,Hoare:2015gda}. There have also been works with the goal of constructing generalisations of Yang-Baxter deformations, see e.g.~\cite{Delduc:2017fib,Hoare:2020mpv}, and in~\cite{Sfondrini:2019smd} it was also possible to show a link to another class of constructions, namely  the $T\bar T$-deformation of~\cite{Smirnov:2016lqw,Cavaglia:2016oda}.

When the original $\sigma$-model has an interpretation in terms of string theory, it is natural to ask whether the deformed models have one too, i.e.~if they give rise to background fields that satisfy the supergravity equations of motion. The conditions under which this happens are by now well under control~\cite{Arutyunov:2015mqj,Borsato:2016ose,Sakamoto:2017cpu,Borsato:2021gma,Borsato:2021vfy}, and the effects of $\alpha'$-corrections have also started to be studied~\cite{Hoare:2019mcc,Borsato:2020bqo}.
In fact, all the above integrable deformations are under intense study also because of the possible applications to generalisations of the AdS/CFT correspondence, see in particular \cite{vanTongeren:2015uha,vanTongeren:2016eeb} and also~\cite{Araujo:2017jkb,Araujo:2017jap} for related work. In the canonical example of the duality between $\mathcal N=4$ super Yang-Mills and the superstring on $AdS_5\times S^5$, the spectral problem admits a reformulation in terms of an integrable model, see~\cite{Beisert:2010jr} for a review. Thanks to that, in the large-$N$ limit, it is possible to write down equations that encode the \emph{exact} information on the two-point functions of the 4-dimensional conformal theory (and on the energies of the dual string configurations) at \emph{finite} values of the 't Hooft coupling. 

While it would be desirable to extend the methods of integrability also to the deformed models and to  compute exact observables also in those cases, progress in this direction has so far been limited only to special cases. In fact, if we restrict the discussion to the family of Yang-Baxter deformations,  it was possible to carry out the program of quantum integrability (that culminates with the Thermodynamic Bethe Ansatz equations or the Quantum Spectral Curve) only in two cases, namely that of the inhomogeneous deformation of $AdS_5\times S^5$~\cite{Arutyunov:2014wdg,Klabbers:2017vtw}, and that of diagonal TsT transformations~\cite{Beisert:2005if,deLeeuw:2012hp}. The latter are obtained by a sequence of T-duality--shift--T-duality~\cite{Frolov:2005ty,Frolov:2005dj} and in fact turn out to be part of the family of hYB deformations~\cite{Osten:2016dvf}. In both cases, a key aspect that allows the extension of the integrability methods is the fact that the deformations preserve the Cartan subalgebra of isometries. On the one hand, this feature allows one to fix the usual BMN light-cone gauge~\cite{Berenstein:2002jq}, that plays an important role in the integrability formulation. On the other hand, one is still able to use  the same number of conserved charges as in the undeformed case to label the states and compute observables.
However, all other hYB deformations --- namely non-diagonal TsT transformations, and all non-abelian deformations like the Jordanian  and almost abelian deformations, see section~\ref{sec:ex} for their definition --- partially break the Cartan subalgebra, and in particular they break the BMN light-cone isometries. As anticipated, this creates major obstacles, even in the conceptually simple cases of non-diagonal TsT deformations~\cite{Guica:2017mtd}.

Here we want to remain in the classical regime and address a direction that  has been largely unexplored, namely the problem of the formulation of the classical spectral curve (CSC) for the hYB deformed models. 
In fact, despite the intensive investigation of  integrable deformations of sigma models, the construction of the CSC has been addressed only for diagonal TsT transformations in~\cite{Frolov:2005ty} and recently for the $\lambda$-deformation  in~\cite{Hollowood:2019ajq}. For more general hYB deformations, a natural strategy is to proceed like in the subclass of TsT transformations \cite{Frolov:2005dj}, where instead of working with a \emph{deformed} $\sigma$-model one exploits an on-shell map to reformulate the dynamics in terms of an \emph{undeformed yet twisted} model. As we will review in section~\ref{sec:osi}, this is simply a consequence of the possibility of identifying the Lax connection of the hYB model with the Lax connection of the undeformed model. When doing this, if we assume that the hYB model satisfies periodic boundary conditions on the worldsheet (which is well motivated for example from the point of view of string theory), the price that we pay is that the field parametrising the undeformed model must instead satisfy twisted boundary conditions. If $\eta$ is the deformation parameter of the hYB model, the twist will depend on $\eta$ and on some conserved charges, a situation which parallels the one in the TsT case~\cite{Frolov:2005dj}. In fact, it was already known since~\cite{Matsumoto:2015jja,Vicedo:2015pna}  that the most generic hYB deformation can be reinterpreted as an undeformed model with twisted boundary conditions.  However, the expression for the twist --- written as a functional of the degrees of freedom of the hYB model --- is in terms of a path-ordered exponential. The non-locality that this  introduces  makes it impractical  to work with the twisted model~\cite{vanTongeren:2018vpb}. The only exceptions are obviously the already known cases of TsT transformations, for which some simplifications (that are consequences of the abelian nature of the algebra related to the twist) reduce the expression to a local one. So far this has created a major obstacle in the formulation of the CSC for a generic hYB deformation, and it has prevented further study of some interesting cases such as the Jordanian and almost abelian deformations.

In this paper we will solve this problem by rewriting the twist of the boundary conditions in a way that is more useful for practical calculations. Instead of writing it in terms of the degrees of freedom of the hYB model, we will write it in terms of those of the twisted model, which is conceptually more natural. The differential equation that we find in this case is non-standard but, remarkably, we are able to solve it in full generality for all hYB deformations. Even more importantly, our solution for the twist is  given by expressions that are \emph{local} in terms of the degrees of freedom of the twisted model. 

The paper is organised as follows. In section~\ref{sec:yb} we review some known facts about the hYB deformation, in particular on the algebraic ingredients that are needed and on its  integrability. We will also make remarks on some known facts about classical integrability that will be important for the  discussion that follows. In section~\ref{sec:osi} we rewrite the hYB deformation as an undeformed yet twisted model, and we present the derivation of our solution for the twist. We also spell out the consequences of the twist at the level of the monodromy matrix. In most of the paper, for simplicity, we deal with deformations of the Principal Chiral Model (PCM), but in section~\ref{sec:sup-cos} we show that our results are easily generalisable  to cosets and supercosets. In section~\ref{sec:ex} we consider different classes of hYB deformations (namely abelian, almost abelian and Jordanian), we construct the twists in each case and discuss their consequences at the level of the CSC. We then finish in section~\ref{sec:concl} with some conclusions. Appendix~\ref{app:CSC} contains material on the CSC of the undeformed models that is useful for section~\ref{sec:ex}. 
In appendix~\ref{app:JordanForm} we review some aspects of non-diagonalizable matrices and their relation to the existence of branch points of the monodromy matrix and the CSC.

\section{The homogeneous Yang-Baxter deformation}\label{sec:yb}

In this section we introduce our conventions, and we review some known facts of the hYB deformation of the PCM and of classical integrability that we will use later. 

The dynamical fields of the  PCM and its hYB deformations will be called $\tilde{g}\in G$ and $g\in G$, respectively. Both take values 
 in a real Lie group $G$ and depend on the worldsheet coordinates $\tau$ and $\sigma$. 
In this paper we will  consider both compact and non-compact groups and, in fact, the only assumption that we need is that the Lie algebra $\alg g $ of $G$ is equipped with a \emph{non-degenerate} symmetric and ad-invariant bilinear form. Even though  {it is} not necessary, for simplicity we will have  a matrix realisation {of} $\alg g$ in mind, and for the bilinear form we will take the one induced by  the trace of products of matrices. 
The action of the PCM is then defined in terms of the left-invariant Maurer-Cartan form $\tilde{J}=\tilde{g}^{-1}d\tilde{g}$
as
\be\label{eq:SPCM}
S_{\text{PCM}}=-T\int d^2\sigma \ \Tr\left(\tilde{J}_+  \tilde{J}_-\right).
\ee
Here $T$ is an overall coupling that will play no role in our discussion, and we are using the light-cone combinations  $\sigma^\pm=(\tau\pm \sigma)/2$ of the worldsheet coordinates, so that $\tilde{J}_\pm=\tilde{g}^{-1}\partial_\pm \tilde{g}$. 
This action is invariant under  global $G_L\times G_R$ transformations
\be
\label{eq:globPCM}
\tilde{g}\to h_L\, \tilde{g}\, h_R^{-1}, \qquad h_L, h_R\in G.
\ee
It gives rise to the equations of motion
\be
\partial_+ \tilde{J}_- +\partial_- \tilde{J}_+=0\,,
\ee
while the currents $\tilde{J}_\pm$ satisfy the Maurer-Cartan identity 
\be
\partial_+ \tilde{J}_- -\partial_- \tilde{J}_+ +[\tilde{J}_+,\tilde{J}_-]=0 \ ,
\ee
off-shell. Classical integrability amounts to the fact that these two sets of equations can be written together as a single zero-curvature condition
\be
\big[\partial_+ +\mathcal{L}_+(z),\partial_- +\mathcal{L}_-(z)\big]=0\,, \qquad \forall z\in{\mathbb C}\, ,
\ee
in terms of the Lax connection
\be
\label{eq:LPCM}
\mathcal{L}_\pm (z)= \frac{\tilde{J}_\pm}{1\mp z},
\ee
which, in addition to $\tau$ and $\sigma$, depends on the auxiliary complex spectral parameter $z$.

To construct the deformed model, one first introduces the linear operator 
\be
\mathcal O\equiv 1-\eta R_g:\alg g\to \alg g \ ,
\ee
{acting on the Lie algebra $\alg g$,} where $\eta$ is a real deformation parameter, so that in the limit $\eta\to 0$ one {will recover} the PCM action. When $\eta\neq 0$, the deformation is driven by the linear operator $R_g:\alg g\to\alg g$ defined as\footnote{Notice that we use a notation such that $O_1O_2$ means the composition of the linear operators $O_i$ on $\alg g$, and that we always assume that the linear operators are acting on the Lie algebra elements that stand on their right. In other words, if $x\in \alg g$ then $O_1O_2x=O_1(O_2(x))$.} $R_g\equiv \AD_g^{-1} R \AD_g$, in terms of a $g$-independent linear operator $R$, and of $\AD_g$, where $\AD_gx=gxg^{-1}$. The action of the hYB deformed PCM is then \cite{Klimcik:2002zj,Klimcik:2008eq}
\be\label{eq:SYB}
S_{\text{YB}}=-T\int d^2\sigma \ \Tr\left(J_+\mathcal O^{-1} J_-\right),
\ee
where
$J_\pm=g^{-1}\partial_\pm g$, and the invertibility of $\mathcal O$ is ensured  at least in a neighbourhood of $\eta=0$. 
{
This action is invariant under the subset of global $G_L\times G_R$ transformations
\be
g\to u_L\, g\, u_R^{-1}, \qquad u_L, u_R\in G.
\label{eq:globYB}
\ee
singled out by the condition
\be
\label{eq:condYB}
\text{Ad}^{-1}_{u_L}\, R\, \text{Ad}_{u_L}=R\,.
\ee
Note that $u_R$ is not restricted and thus the full $G_R$ global symmetry is preserved under the deformation.
}

\subsection{Algebraic ingredients}\label{sec:yb-alg}

The operator $R$ is required to satisfy two algebraic properties such that the hYB model is classically integrable (see section \ref{sec:yb-int}) \cite{Klimcik:2002zj,Klimcik:2008eq}. First, it must be antisymmetric $R^T=-R$, where transposition is implemented via the bilinear form on $\alg g$. In other words, if $x,y\in \alg g$ then the transpose of a generic linear operator $O$ is defined by $\Tr((O^Tx)y)=\Tr(xOy)$,\footnote{{The assumption of non-degeneracy of the bilinear form on $\alg{g}$ ensures that the transpose of a linear operator is uniquely defined. In particular, if $\tilde{O}=0$ holds then $\tilde{O}^T=0$ is implied.}} so that antisymmetry of $R$ reads $\Tr((Rx)y)=-\Tr(xRy)$.
Notice that $(\AD_g)^T=\AD_g^{-1}$.
Second, $R$ must satisfy the classical Yang-Baxter equation (CYBE)\footnote{In the construction of the inhomogeneous YB deformation, $R$ satisfies the \emph{modified} CYBE on $\alg g${, which includes an additional term of the form $-c^2[x,y]$ on the right-hand-side \cite{Klimcik:2008eq}}.} on $\alg g$ \cite{Kawaguchi:2014qwa,Matsumoto:2015jja}
\be\label{eq:cybe}
[Rx,Ry]-R([Rx,y]+[x,Ry])=0,\qquad\qquad
\forall x,y\in \alg g.
\ee

The operator $R$ gives rise to a useful decomposition of $\alg{g}$ as a vector space in terms of the kernel and the dual of the image of $R$. First of all, let us call $\alg f=\text{Im}(R)$ the image of $R$, which is a subalgebra of $\alg g$ as a consequence of the CYBE~\eqref{eq:cybe}. 
Next, let $\alg f^*$ be the dual space of $\alg f$ with respect to the bilinear form on $\alg g$. To be more explicit, choose a basis $\{T_a\,, a=1,\ldots,\text{dim}(\alg g)\}$ of $\alg g$ such that the subset $\{T_i\,, i=1,\ldots,\text{dim}(\alg f)\}$ is a basis of $\alg f$. Then, using that the bilinear form is non-degenerate, we can take $\{T^i=(\kappa^{-1})^{ia}T_a\}$
as a basis of $\alg f^*$, where {$\kappa_{ab}=\Tr(T_aT_b)$} so that $\Tr(T_i T^j)=\delta_i^j$.\footnote{In the definition of $T^i$ it is important to sum over all {$a=1,\ldots,\text{dim}(\alg g)$} since $\kappa$ is invertible on the whole $\alg g$ but it may be degenerate when restricting to $\alg f$, which happens in the case of $\alg g$ non-compact. When $\alg g$ is compact, one can take $\kappa_{mn}\propto \delta_{mn}$ and then $\alg f=\alg f^*$. However, when $\alg g$ is non-compact, the vector spaces of $\alg f$ and $\alg f^*$ do not need to coincide.}
Using now that $R$ is antisymmetric, one can show that $\alg{f}^\ast\cap \text{Ker}(R)=\{0\}$ as follows. Let us write a generic element $x\in \alg f^*$  as $x= \Tr(xT_i)\, T^i$. Then, since $T_i\in\alg{f}$, one can find $y\in\alg{g}$ such that  $T_i=R(y)$ and, thus, $\Tr(xT_i)=-\Tr\big(R(x) y\big)$. Therefore, if $x\in \alg{f}^\ast\cap \text{Ker}(R)$ then $x=0$. Finally, since $R:\alg{g}\to \alg{g}$ is linear, it follows that $\text{dim}(\alg g)=\text{dim}(\alg f) + \text{dim}\,\text{Ker}(R)$, which completes the proof of the decomposition
\be
\label{eq:SplitK}
\alg g = \alg{f}^\ast \oplus \text{Ker}(R).
\ee
In particular, this ensures that the restriction of $R$ to $\alg f^*$, namely $R:\alg f^*\to \alg f$, 
is a bijection and, hence, it is invertible.

Let us denote by $\omega:\alg f\to \alg f^*$ the inverse of $R$. Notice that the fact that we are restricting the operators $\omega$ and~$R$ to $\alg f$ and~$\alg f^*$, respectively, implies that $R\omega=P$ and $\omega R=P^T$, where $P$ and $P^T$ are projectors on $\alg f$ and $\alg f^*$, and they are transpose to each other as the notation suggests.\footnote{
The expansion of a generic element $x\in\alg{g}$ in terms of the basis $T_a, T_b,\ldots$ reads $x=\Tr(xT^a)\, T_a$ where $T^a=(\kappa^{-1})^{ab}T_b$. Then,
\be
Px=\Tr(xT^i)\, T_i, \qquad P^Tx=\Tr(xT_i)\, T^i,
\ee
and thus $\Tr\big(yPx\big)= \Tr\big((P^Ty)x\big)$.}
The fact that $R$ solves the CYBE implies that {$[ \alg{f}, \text{Ker}(R)]\subset\text{Ker}(R)$, and that} $\omega$ satisfies the 2-cocycle condition
\be\label{eq:2cc}
\omega[x,y]=P^T([\omega x,y]+[x,\omega y]),\qquad\qquad\forall x,y\in\alg f,
\ee 
that here takes a non-standard form because of the additional $P^T$.
In operatorial form,~\eqref{eq:2cc} may be equivalently written as
\be
\omega \Ad_xP=P^T(\Ad_{\omega x}+\Ad_x\omega)P,\qquad\qquad\forall x\in\alg f,
\ee
{where $\Ad_xy=[x,y]$.}

Writing $RT^i = R^{ij}T_j$ with $R^{ij}=-R^{ji}$, we can actually represent $R$ as an element  $r\in \alg f\wedge\alg f$ 
\be\label{eq:rTT}
r=- R^{ij}T_i\wedge T_j,
\ee
so that $Rx=\Tr_2(r(1\otimes x))=- R^{ij}\Tr(T_jx)T_i$, where $\Tr_2$ denotes the trace in the second space of the tensor product.  
 Then the CYBE takes the more familiar form 
\be
[r_{12},r_{13}]+[r_{12},r_{23}]+[r_{13},r_{23}]=0,
\ee
where $r_{ab}$ denotes $r$ in the $a$ and $b$ spaces of the triple tensor product. Similarly, one can define $\hat\omega:\alg f\wedge\alg f\to \mathbb R$ as $\hat\omega(x,y)=\Tr(x\omega y)$, so that the 2-cocycle condition reads
\be
\hat\omega(x,[y,z])+\hat\omega(y,[z,x])+\hat\omega(z,[x,y])=0.
\ee
{The simplest} solutions of the CYBE are obtained by choosing $\alg f$  to be abelian. However, there also exist interesting solutions of the CYBE where $\alg f$ is non-abelian, and we shall discuss some examples in section~\ref{sec:ex}.  Being linear, the 2-cocycle condition is easier to solve than the CYBE. In general, the solutions to the CYBE on $\alg g$ are classified by the quasi-Frobenius subalgebras $\alg f\subset \alg g$ \cite{Stolin1991a,STOLIN1999285,Gerstenhaber1997}. An algebra $\alg f$ is said to be quasi-Frobenius if it admits an invertible 2-cocycle. It is Frobenius if the cocycle is also a coboundary, meaning that there exists an $x\in\alg f$ such that $\omega =\Ad_x$. For more details on the relation between CYBE and quasi-Frobenius algebras we refer the reader to~\cite{STOLIN1999285}.

\subsection{Classical integrability}\label{sec:yb-int}
Computing the variation of the YB action under $g\to g+\delta g$ one finds
\be\label{eq:deltaS}
\delta S_{\text{YB}}=T\int d^2\sigma\ \Tr\left[g^{-1}\delta g \left(\partial_+A_-+\partial_-A_+\right)\right]+b,
\ee
where
\be\label{eq:Apm}
A_-=\mathcal O^{-1}J_-=\frac{1}{1-\eta R_g}(g^{-1}\partial_- g),\qquad
A_+=\mathcal O^{-T}J_+=\frac{1}{1+\eta R_g}(g^{-1}\partial_+ g),\qquad
\ee
and the boundary term is
\be\label{eq:bound}
b=-T\int d^2\sigma \Tr\left[\partial_+\left(g^{-1}\delta gA_-\right)+\partial_-\left(g^{-1}\delta gA_+\right)\right].
\ee
To derive $\delta S_{\text{YB}}$ in~\eqref{eq:deltaS} we have used $\delta \AD_g=\AD_g\Ad_{g^{-1}\delta g}$ and $\delta R_g=[R_g,\Ad_{g^{-1}\delta g}]$. Notice, in particular, that  no property of $R$ (neither antisymmetry nor CYBE) is  needed.
The boundary term vanishes upon taking periodic boundary conditions $g(\tau,\sigma+2\pi)=g(\tau,\sigma)$, which is the choice we make for the hYB model. Therefore, the equations of motion of the deformed model are 
\be
\partial_+A_-+\partial_-A_+=0,
\ee
which are formally equivalent to those of the PCM ($\partial_+\tilde{J}_-+\partial_-\tilde{J}_+=0$) after the replacement $\tilde{J}\to A$. In the PCM, $\tilde{J}$ satisfies the Maurer-Cartan identity \emph{off-shell}. In the hYB deformation, $J=g^{-1}\partial g$ also satisfies the Maurer-Cartan identity, and rewriting it in terms of $A$ one finds
\be
\begin{aligned}
0&= \partial_+J_--\partial_-J_++[J_+,J_-]\\
&= \partial_+A_--\partial_-A_++[A_+,A_-]\\
&-\eta(R_g\partial_+A_--R_g^T\partial_-A_++(R_g+R_g^T)[A_+,A_-])\\
&-\eta^2([R_g^TA_+,R_gA_-]-R_g[R_g^TA_+,A_-]-R_g^T[A_+,R_gA_-]).
\end{aligned}
\ee
Upon requiring antisymmetry of $R$, the term at  order $\eta$ is proportional to the equations of motion. It therefore vanishes on-shell, albeit  not off-shell. The term at order $\eta^2$ vanishes if, together with antisymmetry, we also demand that $R$ (and therefore $R_g$)  satisfies the CYBE.

To conclude, when $R$ is antisymmetric and solves the CYBE, $A_\pm$ satisfy the following \emph{on-shell} equations
\be
\partial_+A_-+\partial_-A_+=0,\qquad\qquad
\partial_+A_--\partial_-A_++[A_+,A_-]=0.
\ee
These are formally the same equations as those satisfied by $\tilde{J}$ in the PCM case, and this is enough to conclude that the YB deformation preserves classical integrability. 
Namely, we may introduce the Lax connection
\be\label{eq:L}
\mathcal L_\pm(z)=\frac{A_\pm}{1\mp z},
\ee
so that the above on-shell equations are equivalent to the flatness of the Lax connection
\be
\partial_+\mathcal L_-(z)-\partial_-\mathcal L_+(z)+[\mathcal L_+(z),\mathcal L_-(z)]=0\,,\qquad \forall z\in{\mathbb C}\,  .
\ee
Note that, in addition we can write them as
\begin{equation}
\partial_\pm A_\mp = \mp \frac12 [A_+ , A_- ] ,
\end{equation}
which implies that
\begin{equation} \label{eq:LocalCCL}
\partial_\pm \Tr (A_\mp^n) = 0,
\end{equation}
for all $n\in \mathbb{Z}$. This provides an infinite set of local conserved currents $\Tr (A_\pm^n)$ as discussed in \cite{Evans:1999mj,Evans:2000hx,Evans:2000qx}, and reflects the fact that the theory is classically conformal invariant.

At this point {we will continue with the Lax perspective and} use the standard machinery of integrability that we briefly recap in the following (see e.g.~also \cite{Babelon:2003qtg}). From the Lax connection one may define the monodromy matrix in terms of a path-ordered exponential
\be\label{eq:Omega}
\Omega(z,\tau) = \mathcal P\exp\left(-\int_0^{2\pi}d\sigma'\mathcal L_\sigma(z,\tau,\sigma')\right),
\ee
{which satisfies
\be
\partial_\tau\Omega(z,\tau) =-\mathcal L_\tau(z,\tau,2\pi)\Omega(z,\tau)+\Omega(z,\tau)\mathcal L_\tau(z,\tau,0).
\ee
Therefore, when} $\mathcal L_\tau$ is periodic in $\sigma$, which means that $\mathcal L_\tau(z,\tau,2\pi)=\mathcal L_\tau(z,\tau,0)$ like in our case, the eigenvalues {$\lambda(z)$} of the monodromy matrix are constant, and by expanding $\lambda(z)$ in powers of $z$ one obtains an infinite tower of conserved quantities.

The information on the eigenvalues of the monodromy matrix is encoded in the classical spectral curve (CSC), which is {defined by} the characteristic equation\footnote{See e.g.~\cite{Vicedo:2008ryn} for a more detailed discussion on the nature of the curve.}
\be\label{eq:det}
\det\left[\lambda(z)\mathbf 1-\Omega(z,\tau)\right]=0.
\ee
Taking {$\Omega(z,\tau)$} to be a $M\times M$ matrix, we will have eigenvalues {$\lambda_A(z), A=1,\ldots, M$}. In general {$\Omega(z,\tau)$} does not need to be diagonalisable, and at most we can hope to put it into Jordan form. Later we will see examples of this kind.
{Invoking a well-known theorem by Poincar\'e \cite{Babelon:2003qtg}, the monodromy matrix $\Omega(z)$ is analytical in $z$ except at the poles of the Lax connection (in our case $z=\pm 1$) where it has essential singularities.  This makes the curve obtained by~\eqref{eq:det}   highly singular. }
To deal with the  essential singularities,  it is useful to introduce the parameterisation 
\be \label{eq:DefQM}
\lambda_A(z)=e^{ip_A(z)},
\ee
in terms of quasimomenta $p_A(z)$, which we can think of as the components of an element of the Cartan subalgebra of $\alg{g}$. In fact, the {quasimomenta} have simple poles  at $z=\pm 1$.\footnote{ \label{f:Omdiag}Let us denote by $v_\pm$ the matrices diagonalising $ A_\pm$,
\be
 A_\pm^{\text{diag}}(\tau,\sigma)=v_\pm(\tau,\sigma)\  A_\pm(\tau,\sigma)\ v_\pm^{-1}(\tau,\sigma).
\ee
One can show that around $z=\pm 1$ it is possible to construct gauge transformations $u_{\pm}(z,\tau,\sigma)$ such that 
\emph{(i)}
$u_\pm(z,\tau,\sigma)=v_\pm(\tau,\sigma)+\mathcal O(z\mp 1)$,
\emph{(ii)}
$u_\pm(z,\tau,\sigma)$ are periodic in $\sigma$,
\emph{(iii)}
$\mathcal L_\sigma^{\text{diag},\pm}= u_{\pm} \mathcal L_\sigma u_{\pm}^{-1} -\partial_\sigma u_{\pm}u_{\pm}^{-1}$ are diagonal to all orders in the $z$-expansion.
See e.g.~\cite{Beisert:2004ag} for a discussion on this.
From~\eqref{eq:L}  we then have
\be
\mathcal L_\sigma^{\text{diag},\pm}=\pm \frac12 \frac{ A_\pm^{\text{diag}}}{1\mp z}+\mathcal O\big((1\mp z)^{n\geq0}\big).
\ee
Because $\mathcal L_\sigma^{\text{diag},\pm}$ is diagonal, one can drop the path-ordering in the exponential defining the monodromy matrix. Around $z=\pm 1$ it takes the form
\be\label{eq:sing-Omega}
u_\pm(z,0)\ \Omega(z)\ u_\pm(z,0)^{-1}=\exp\left(\mp \frac12\int_0^{2\pi} d\sigma \frac{A_\pm^{\text{diag}}}{1\mp z}+\mathcal O\big((1\mp z)^{n\geq0})\right),
\ee
where we used $u_\pm(z,2\pi)=u_\pm(z,0)$ and we omitted the $\tau$-dependence for brevity.
Note that  from \eqref{eq:LocalCCL} we know that $A_\pm^{\text{diag} }= A_\pm^{\text{diag}} (\tau \pm \sigma)$.
Together with the fact that $A_\pm^{\text{diag} }$ is periodic this implies that the object
\begin{equation} \label{eq:defm}
q^\pm \equiv \frac{i}{2}\int^{2\pi}_0 d\sigma A_\pm^{\text{diag}} (\tau \pm \sigma) =\text{diag}(q^\pm_1,\ldots,q^\pm_M) \ ,
\end{equation}
is a constant diagonal matrix, so that 
\be \label{eq:ResiduesQM0}
p_A(z)\approx \pm \frac{q^\pm_A}{1\mp z}\quad \text{at}\quad  z\approx \pm 1.
\ee
}

The characteristic equation~\eqref{eq:det} is a polynomial equation  for the function $\lambda(z)$ and, in general, it gives rise to branch points for  $\lambda(z)$ where two or more eigenvalues become degenerate (see appendix~\ref{app:JordanForm}). The number and the location of the branch points is  data that depends on the particular solution of the equations of motion, and one may actually use this data as a way to classify families of such solutions. We therefore have $M$ Riemann sheets, one for each of the eigenvalues, glued together along the cuts specified by the branch points.   
Denoting by $\mathcal C_{(A,B)}$ the branch cut (or more generally the union of all branch cuts) that joins the Riemann sheets
corresponding to $\lambda_{A}(z)$ and $\lambda_{B}(z)$, the relation becomes
\be\label{eq:ppn}
p_A(z+i\epsilon)- p_B(z-i\epsilon)=2\pi n_{(A,B)},\qquad \text{for } \ z\in \mathcal C_{(A,B)} ,
\ee 
in terms of the quasimomenta, for some integers $n_{(A,B)}$.
We then have a collection of Riemann-Hilbert problems. 

When discussing the analytic properties of $p_A(z)$ it is important to notice that, as anticipated in~\eqref{eq:ResiduesQM0}, it only has simple pole singularities {with constant residues}
\be \label{eq:ResiduesQM}
p_A(z)\approx \pm \frac{q^\pm_A}{1\mp z},\qquad z\approx \pm 1  .
\ee
As in~\cite{Kazakov:2004qf,Beisert:2004ag,Beisert:2005bm}, we therefore define the (diagonal matrix) resolvent
\be \label{eq:defResolvent}
G(z)\equiv p(z)+\frac{q^-}{1+ z}-\frac{q^+}{1- z},
\ee
where these singularities are removed. The resolvent is thus analytic everywhere in the complex plane with cuts. As a consequence of this, we can represent $G(z)$ in terms of ``densities'' $\rho(z)$ that are functions with support only on the cuts $\mathcal C$
\be
G(z)=\int_{\mathcal C}dy\frac{\rho(y)}{z-y},
\ee
where  $\mathcal C$ is understood as the union of all the cuts. The so-called ``finite-gap solutions'' are a class of solutions of the classical spectral problem, where only a finite number $K$ of square-root cuts is allowed \cite{Novikov:1984id}.\footnote{In general, one may have higher-order cuts that join more than two sheets, where more than two eigenvalues coincide. To the best of our knowledge, finite-gap equations refer to the case of square-root cuts only.}  In this context the equations~\eqref{eq:ppn} go under the name of ``finite-gap equations'' {or ``classical Bethe equations''}, and are usually written as integral equations for the densities $\rho(z)$ \cite{Kazakov:2004qf}. 

In the case of the undeformed and periodic PCM, the quasimomentum turns out to have the following ``asymptotics'' around $z=0$ and $z=\infty$ (see also section \ref{ss:mon})
\begin{equation}\label{eq:as-p}
\begin{alignedat}{2}
& p(z)\approx p(0) + z \ p'(0) +\mathcal O(z^2),\qquad &&\text{ around }z=0,\\
& p(z)\approx \frac{ p_{\infty}}{z}+\mathcal O(z^{-2})\qquad &&\text{ around }z=\infty,
\end{alignedat}
\end{equation}
where $p_{\infty}$ is constant. Actually, as a consequence of the periodic boundary conditions, $\lambda(0)=1$ and therefore, after taking the log, $p(0)=2\pi m$ with $m$ an integer. This introduces a multi-valuedness of $p(z)$ that can be overcome by considering $dp(z)$ instead. The same ambiguity is obviously there also for the deformed/twisted models, where the value $p(0)=2\pi m$ will be shifted by constants related to the particular twist to be discussed in the next section. Knowing the properties of the quasimomentum, $dp$ turns out to be an abelian differential of the second kind (i.e.~locally it is $dp=f(z)dz$ where $f(z)$ is meromorphic with vanishing residues).
The idea of~\cite{Dorey:2006zj} (see also~\cite{Vicedo:2008ryn}) is to use a construction due to Krichever and Phong~\cite{Krichever:1996ut} where the authors consider the moduli space of all Riemann surfaces characterised by the following data:
\begin{itemize}
\item The genus $g$ of the Riemann surface. In our case it is given by\footnote{The relation $g=K-(M-1)$ is under the assumption that the collection of sheets describes a connected surface, or in other words that at least one sheet is connected by one cut to all other $M-1$ sheets. In this minimal case we have, for example, $M-1$ Riemann spheres connected to a single Riemann sphere, and the genus is 0. The addition of one cut at a time increases the genus by one unit.} $g=K-(M-1)$ where $K$ is the number of cuts and $M$ the number of sheets. 
\item Punctures $P_{\hat\alpha}$ (with $\hat\alpha=1,\ldots ,\hat N$) on the Riemann surface that correspond to poles of at least one of two abelian integrals denoted by $E$ and $Q$.
The maximum order of the pole in each case is denoted with $n^{(E)}_{\hat\alpha}$ and $n^{(Q)}_{\hat\alpha}$.
\end{itemize}
In~\cite{Krichever:1996ut} it is shown how to construct a full set of independent local coordinates to parameterise the aforementioned moduli space $\mathcal M_g(\{n^{(E)}_{\hat\alpha}\},\{n^{(Q)}_{\hat\alpha}\})$, whose complex dimension turns out to be 
\be
\dim_{\mathbb C} \mathcal M_g\left(\{n^{(E)}_{\hat\alpha}\},\{n^{(Q)}_{\hat\alpha}\}\right) = \sum_{\hat\alpha=1}^{\hat N}\left(n^{(E)}_{\hat\alpha}+n^{(Q)}_{\hat\alpha}\right) +3 (\hat N-1)+5g.
\ee
When the quasimomentum has asymptotics as in~\eqref{eq:as-p}, one can set~\cite{Dorey:2006zj}
\be
E=p,\qquad\qquad Q=z+1/z,
\ee
in the construction of~\cite{Krichever:1996ut}, so that there are four punctures in each sheet, located  at $z=+1,-1,0,\infty$.
With this choice, some of the coordinates of~\cite{Krichever:1996ut} turn out to be model-independent constants (with the choice of~\cite{Dorey:2006zj} they only take the values $0,+2,-2$). In other words, the Riemann surfaces describing the CSC of the integrable models that we want to study all belong to a common subspace of the full moduli space of~\cite{Krichever:1996ut}, characterised by that choice of coordinates. The remaining coordinates turn out to depend precisely on the following data
\begin{itemize}
\item $p(0)$, the value of the quasimomentum at $z=0$,
\item $p'(0)$, the value at $\mathcal O(z)$ of the expansion of the quasimomentum around $z=0$.
\item $p_\infty$, the constant multiplying $1/z$ when expanding around $z=\infty$,
\item $n_{(A,B)}$, the collection of  integers allowed at each collection of cuts $\mathcal C_{(A,B)}$. One may introduce a canonical set of $a$- and $b$-cycles (that we will label with the index $\mu=1,\ldots,g$) for the Riemann surface, such that all $a$-periods of $dp$ vanish ($\oint_{a_\mu}dp=0$,  $\forall \mu=1,\ldots,g$)~\cite{Kazakov:2004qf,Dorey:2006zj}. Then the integers $n_\mu$ (now labelled with the new index $\mu$ rather than $(A,B)$) are related to the $b$-periods of $dp$ as $\oint_{b_\mu}dp=2\pi n_\mu$ with $\mu=1,\ldots,g$,
\item and $S_\mu$, the ``filling fractions'' defined by the $a$-periods of $Qdp$, namely\footnote{In the context of the AdS/CFT correspondence, it is preferable to normalise the filling fractions in terms of the 't Hooft coupling and other numerical constants, such that on the gauge theory side they take discrete values and can be interpreted as the number of Bethe roots condensing on the cuts, see e.g.~\cite{Beisert:2005bm}. We do not pay attention to this normalisation in this paper.} $S_\mu=\oint_{a_\mu}Qdp=-\oint_{a_\mu}pdQ$.  The filling fractions can be equivalently written  in terms of the twist function of the Hamiltonian formulation~\cite{Delduc:2012qb},
which in this case is $\varphi(z)=1-1/z^2$ and  is related to $Q$ as $dQ=\varphi(z) dz$.\footnote{For the symmetric space sigma model, it is $\varphi(z)=\phi(z^2)/z$ with $\phi(z^2)= z^2/(1-z^2)^2$, and for the semisymmetric space sigma model $\varphi(z)=\phi(z^4)/z$~\cite{Delduc:2012vq} (see also~\cite{Hollowood:2019ajq}). }
\end{itemize}
Therefore, the knowledge of the above data is enough to uniquely identify the Riemann surface in the full moduli space $\mathcal M_g(\{n^{(E)}_{\hat\alpha}\},\{n^{(Q)}_{\hat\alpha}\})$ and, at least in principle, reconstruct from this data the whole Riemman surface and quasimomentum. Because the proof is not constructive, more work is needed to actually reconstruct $p(z)$, and we will do this explicitly for  a single cut in appendix~\ref{app:CSC} and section~\ref{sec:ex}. Nevertheless, when the quasimomentum has the asymptotics as in~\eqref{eq:as-p}, we will appeal to the construction of~\cite{Dorey:2006zj,Vicedo:2008ryn,Krichever:1996ut} to argue that the data at our disposal is enough to describe the full CSC of the integrable models.

\section{On-shell equivalence to a twisted undeformed model}\label{sec:osi}

The fact that the Lax connection of a hYB model has the same form as that of a PCM permits to identify the two models \emph{on-shell} so that the infinite tower of conserved quantities is preserved.
This kind of  identification is familiar in the case of (abelian, non-abelian or more generically Poisson-Lie) T-duality, where the original and the dual models have equations of motion on the worldsheet that can be related by the duality map. 
In the covariant approach of~\cite{Witten:1986qs}, where the phase space is identified with the space of solutions to the equations of motion, one may therefore expect that all these theories  share the same phase space. 
This conclusion is correct only if we leave the choice of boundary conditions free. But we will be interested in solutions of the hYB  that satisfy periodic boundary conditions and, as we shall review, the on-shell identification will map them to solutions of a PCM that satisfy \emph{twisted} boundary conditions.
Therefore, to identify the phase space and the spectrum of the periodic hYB model, one should look at a new and twisted (rather than periodic) PCM. Let us first briefly repeat how this is discussed in \cite{Matsumoto:2015jja,Vicedo:2015pna,vanTongeren:2018vpb}.

Let us consider a PCM (with no deformation) parameterised by a group element $\tilde g\in G$. It  admits a Lax connection of  the familiar form~\eqref{eq:LPCM}.
The on-shell identification consists in demanding that, on the solutions to the equations of motion, the Lax connections of the PCM and of the hYB coincide. This amounts to the equality
\be\label{eq:osi}
\tilde J_\pm=A_\pm,
\ee
where $A_\pm$ of the hYB model was defined in~\eqref{eq:Apm}. The condition above is a set of partial differential equations that relate two on-shell field configurations $\tilde{g}$ and $g$. It is convenient to introduce the so-called ``twist field'' $\mathcal F\in G$ such that \cite{Matsumoto:2015jja,Vicedo:2015pna,vanTongeren:2018vpb} 
\be\label{eq:gFgt}
\mathcal F=g\tilde g^{-1}.
\ee
Consistency with~\eqref{eq:osi} implies that
\be\label{eq:dF1}
\partial_\pm\mathcal F= V_\pm\mathcal F,\qquad\qquad
V_\pm=\pm\eta R\left(gA_\pm g^{-1}\right),
\ee
which can be solved in the usual way by
\be\label{eq:solFpe}
\mathcal F(\tau,\sigma)=\mathcal P\exp\left(\int_{(0,0)}^{(\tau,\sigma)} d\sigma'{}^\alpha V_\alpha(\tau',\sigma')\right)\mathcal F(0,0),\qquad
\mathcal F(0,0)=g(0,0)\tilde g^{-1}(0,0).
\ee
Furthermore, equation~\eqref{eq:gFgt} implies that, if we impose periodic boundary conditions on the field $g$ of the hYB model, then the field $\tilde g$ of the PCM must in general have \emph{twisted} boundary conditions, since
\be
\tilde g(2\pi)=\mathcal F^{-1}(2\pi)g(2\pi)=\mathcal F^{-1}(2\pi)g(0)=\mathcal F^{-1}(2\pi)\mathcal F(0)\tilde g(0)\equiv W \tilde g(0),
\ee
where the expressions are understood as evaluated at fixed $\tau$. We will call $W$ the ``twist'', such that
\be\label{eq:tbc}
\tilde g(2\pi)=W\tilde g(0),\qquad
W=\mathcal F^{-1}(2\pi)\mathcal F(0).
\ee
In general $W\neq 1$, although it reduces to 1 when sending $\eta\to 0$. 
Notice that, even though the PCM action is independent of the deformation parameter $\eta$, the twisted field configuration $\tilde{g}$ does depend on~$\eta$.

\subsection{A new solution for the twist}\label{sec:twist}

In~\eqref{eq:solFpe}, $g(0,0)$ and $\tilde{g}(0,0)$ are integration constants. By a simple field redefinition, for example on $\tilde g$, we can always choose them to be equal, $\tilde{g}(0,0)=g(0,0)$,  so that $\mathcal F(0,0)=1$.  With this choice, and calling $F$ the Lie group of $\alg f=\text{Im}(R)$, we have from \eqref{eq:dF1} that the twist field $\mathcal F$ and consequently the twist $W$ belong to  the group   $F$.  In the following we will always assume this, which will prove useful to simplify the calculations.

Both equation~\eqref{eq:dF1} and the initial conditions $ \mathcal F(0,0)=1$ are invariant under the transformations
\be
\label{eq:trans}
\mathcal F\to u_L \mathcal F u_L^{-1}, \qquad g\to u_L\, g\, u_R^{-1},
\ee
where $u_L, u_R\in G$, and $u_L$ satisfies the condition~\eqref{eq:condYB}, 
$\text{Ad}^{-1}_{u_L}\, R\, \text{Ad}_{u_L}=R$, which  preserves   $\mathcal F\in F$. Moreover, using the definition of the twist field in~\eqref{eq:gFgt}, they lead to
\be\label{eq:new-tr-gt}
\tilde{g}\to u_L\, \tilde{g}\, u_R^{-1},
\ee
which shows that the transformations~\eqref{eq:trans} are a subset of the global transformations of the PCM and the hYB deformation given by~\eqref{eq:globPCM} and~\eqref{eq:globYB}. 
Note that the twist $W$ then transforms in the same way as ${\cal F}$, i.e.~$W\to u_L  W u_L^{-1}$, again consistent with $W\in 
F$. 

Let us stress that $W$ is constant,  which means that it does not depend on the worldsheet coordinates $\tau,\sigma$. 
This easily follows from its definition in~\eqref{eq:tbc} and the fact that $V_\pm$, given by~\eqref{eq:dF1}, satisfies periodic boundary conditions,
\be
\partial_\pm W = -\mathcal F^{-1}(2\pi)\left(V_\pm(0)-V_\pm(2\pi) \right)
\mathcal F(0)=0\,.
\ee
In other words, the twist is  one of the conserved quantities of the model and, importantly, this means that its value will depend on the dynamical solution, so that different field configurations may have twists with different values. This mechanism is in fact already familiar from the study of TsT transformations~\cite{Frolov:2005ty,Frolov:2005dj}.
 Another observation is that  a PCM with a field that satisfies twisted boundary conditions as in~\eqref{eq:tbc} can be defined by an action of the standard form~\eqref{eq:SPCM}, with no need to add more boundary terms.\footnote{In fact, when varying the action by a $\delta\tilde g$ that satisfies $\delta\tilde g(2\pi)=W\delta\tilde g(0)$, as appropriate in this case, one immediately finds the desired  equations of motion $\partial_+\tilde J_-+\partial_-\tilde J_+=0$ because the boundary terms still vanish (cf.~\eqref{eq:bound} with $g\to\tilde g$ and $A\to \tilde J$).}

Eq.~\eqref{eq:solFpe} provides an expression for the twist field $\mathcal F$ as a non-local function of the hYB field~$g$ that is however not useful for practical purposes. In particular, the complicated non-local expressions coming from the path-ordered exponential make it difficult to obtain a satisfactory description of the CSC, see~\cite{vanTongeren:2018vpb}. Here we solve this problem by rewriting the twist field $\mathcal F$ in terms of $\tilde g$ rather than $g$. This is in fact natural, since it is the field $\tilde g$ that is twisted by $W$. We then rewrite equation~\eqref{eq:dF1} in terms of $\mathcal F$ and $\tilde g$ only, and obtain\footnote{In the rest of this section, we prefer to use the Hodge dual notation, such that $*\partial_\pm=\pm\partial_\pm$. Then, more explicitly, equation~\eqref{eq:dF2} reads 
$\partial_\pm\mathcal F\mathcal F^{-1} = \pm\eta R\big(\mathcal{F}\, \partial_\pm\tilde g\tilde g^{-1}\, \mathcal{F}^{-1} \big)$.}
\be\label{eq:dF2}
d\mathcal F\mathcal F^{-1} = \eta R\AD_{\mathcal F}(*d\tilde g\tilde g^{-1}).
\ee
This equation is perhaps non-standard compared to~\eqref{eq:dF1} but, importantly,  {we will see that} we are able to find a solution for $\mathcal F$ and, even more remarkably, that this solution  is actually a \emph{local} expression in terms of $\tilde g$.
First, let us multiply~\eqref{eq:dF2} by $\omega$ from the left, where $\omega$ is the 2-cocycle reviewed in section~\ref{sec:yb-alg}.
Then, the equation becomes
\be\label{eq:dF2a}
\eta^{-1} \omega_{\mathcal{F}}(\mathcal F^{-1} d\mathcal F)=\AD_{\mathcal F}^{-1}P^T\AD_{\mathcal F}(*d\tilde g\tilde g^{-1}),\qquad
 \omega_{\mathcal{F}}\equiv  \AD_{\mathcal F}^{-1} \omega \AD_{\mathcal F}.
\ee
If we now project both sides of~\eqref{eq:dF2a} with $P^T$ and we use the identity\footnote{This identity is the transpose of $\AD_{\mathcal F}^{-1}P\AD_{\mathcal F}P=P$ which, since $\mathcal{F}\in F$, is trivially satisfied.} $P^T\AD_{\mathcal F}^{-1}P^T\AD_{\mathcal F}=P^T$, we obtain the simpler equation
\be\label{eq:dF3}
\eta^{-1} P^T \omega_{\mathcal{F}}(\mathcal F^{-1} d\mathcal F)=P^T(*d\tilde g\tilde g^{-1}).
\ee
The nice thing about the ``projected'' equation~\eqref{eq:dF3} is that we managed to separate variables: all $\mathcal F$-dependence is on the left-hand-side and all $\tilde g$-dependence on the right-hand-side. 

In a moment we will see how to solve~\eqref{eq:dF3}. But, before doing that, let us show  that the projected equation~\eqref{eq:dF3} actually implies~\eqref{eq:dF2a}. Consider the right-hand-side of~\eqref{eq:dF2a} and insert the identity $1=(1-P^T)+P^T$, to obtain
\be
\AD_{\mathcal F}^{-1}P^T\AD_{\mathcal F}(1-P^T)(*d\tilde g\tilde g^{-1})+\AD_{\mathcal F}^{-1}P^T\AD_{\mathcal F}P^T(*d\tilde g\tilde g^{-1}).
\ee
The first term above  vanishes because of the identity $P^T\AD_{\mathcal F}(1-P^T)=0$,\footnote{Taking into account~\eqref{eq:SplitK}, $1-P^T$ projects on $\text{Ker}(R)$ and, thus, this identity follows from the fact that $\mathcal{F}\in F$ and $[\alg{f},\text{Ker}(R)]\subset \text{Ker}(R)$.} and in the second term  we can use~\eqref{eq:dF3}. After  using the definition of $\omega_{\mathcal{F}}$, we immediately find the left-hand-side of~\eqref{eq:dF2a}. 

In the following it will therefore be enough to solve the simpler equation~\eqref{eq:dF3}. To do that, let us define $Y\in\alg f^*$  such that
\be\label{eq:red-y-F}
Y\equiv\frac{1}{\eta} P^T\frac{1-\AD_{\mathcal F}^{-1}}{\log\AD_{\mathcal F}}\omega\log\mathcal F=\frac{1}{\eta} P^T\frac{1-e^{-\Ad_{RX}}}{\Ad_{RX}}X,
\ee
where we rewrote the twist field  in terms of $X\in \alg f^*$ as
\be
\mathcal F=\exp(RX).
\ee
Given that $\mathcal F\in F$ and that $R:\alg f^*\to \alg f$ is a bijection, there is in fact a one to one map between $\mathcal F$ and $X$. 
Equation~\eqref{eq:red-y-F} is a map between $\mathcal F$ (or $X$) and the new field $Y$ that is in  general highly non-linear. Nevertheless,  notice that it is perfectly local, and because it is just a diffeomorphism in target space it is in fact local in the strong sense.
Equation~\eqref{eq:red-y-F} should be understood as a formal series defined by the series of the exponential, i.e.
\be\label{eq:red-y-X}
\begin{aligned}
Y&=\frac{1}{\eta}P^T\ \sum_0^\infty\frac{(-1)^n}{(n+1)!}\Ad_{RX}^nX\\
&=\frac{1}{\eta}P^T\left(X-\tfrac12 [RX,X]+\tfrac16 [RX,[RX,X]]+\ldots\right).
\end{aligned}
\ee
Introducing $Y$ turns out to be the key to the solution, because from its definition it follows that~\cite{Borsato:2016pas,Borsato:2017qsx}
\be\label{eq:dydf}
dY=\frac{1}{\eta}P^T\omega_{\mathcal F}(\mathcal F^{-1} d\mathcal F).
\ee
This formula may be obtained by noticing that $\omega$ formally acts as a derivative, or it can be proved by brute force.
From~\eqref{eq:dF3}  it then follows that
\be\label{eq:y-tg}
dY=P^T(*d\tilde g\tilde g^{-1}).
\ee
Notice that we must have $d^2Y=0$, and this is in fact consistent with the above equation because $d(*d\tilde g\tilde g^{-1})=0$ by the equations of motion of the twisted PCM.
The solution to~\eqref{eq:y-tg} is 
\be
Y(\tau,\sigma)=\int_{\gamma(\tau,\sigma)}P^T(*d\tilde g\tilde g^{-1})+\text{const},
\ee
for some path $\gamma(\tau,\sigma)$ or, more explicitly,
\be\label{eq:soly}
Y(\tau,\sigma) = P^T\left(\int_0^\sigma d\sigma'(\partial_\tau\tilde g\tilde g^{-1}) (\tau,\sigma')+\int_0^\tau d\tau'(\partial_\sigma\tilde g\tilde g^{-1} )(\tau',0)\right)+Y(0,0),
\ee
where we have chosen a particular path. 

To conclude, our solutions for the twist field and the twist are
\be
\mathcal F=e^{RX},\quad\qquad
W=e^{-RX(2\pi)}e^{RX(0)},
\ee
where $X$ is related to $Y$ as in~\eqref{eq:red-y-F} and $Y$ is related to $\tilde g$ as in~\eqref{eq:soly}. The only job left to do is to invert~\eqref{eq:red-y-F} and write a closed expression for $X$ in terms of $Y$. {In general it is} difficult to find such an explicit closed-form formula for an unspecified $R$. However, once we specify $\alg g,\alg f$ and $R$,  this becomes  a simple task because $\alg f$ is finite-dimensional and the calculations amount to simple operations in linear algebra. We will discuss some examples of this procedure in the next section.

Notice that  we can always write the twist as
\be\label{eq:WRQ}
W=\exp(-\eta R\mathcal Q),
\ee
for some $\mathcal Q\in \alg f^*$ which is a conserved charge because $W$ is constant. We remind that, in the undeformed and periodic PCM, $d\tilde g\tilde g^{-1}$ is in fact the Noether current for left transformations, whose integral gives the conserved charges corresponding to the left $G$ isometries (the transformations corresponding to $h_L$ in~\eqref{eq:globPCM}). In the presence of the twist induced by the hYB deformation, some of these isometries are broken, but $d\tilde g\tilde g^{-1}$ is still playing a role (albeit non-trivial) in identifying $\mathcal F, W$ and obviously the conserved charge $\mathcal Q$.

As we have already remarked, under the symmetry transformations~\eqref{eq:new-tr-gt}  of the PCM  field, the twist transforms as $W\to u_L W u_L^{-1}$, with $\AD_{u_L}R=R\AD_{u_L}$. This is enough to conclude that in general we should consider \emph{equivalent classes} of twists, related by the adjoint action by $u_L$. One may in fact exploit this freedom to try to simplify the expression for the twist. For example,  in some cases it may be possible to diagonalise $W$, so that it only takes values in the Cartan subgroup of $F$. In the most general case, it will be possible to write $W$ in Jordan block form, and some examples of non-diagonalisable twists will indeed appear later.

\subsection{The monodromy matrix at special points} \label{ss:mon}
As reviewed in section~\ref{sec:yb-int}, the construction of the CSC requires the knowledge of the asymptotic values of the quasimomenta at $z=0$ and $z=\infty$. We therefore need to discuss the asymptotic values of the monodromy matrix around those points. From now on, we will always be working with the twisted PCM parameterised by the field $\tilde g$.

Let us start witht the expansion around $z=\infty$.  From~\eqref{eq:LPCM}, it follows that 
\be
\mathcal L_\sigma= -\frac{\tilde J_\tau}{z}+\mathcal O(z^{-2}),\qquad\qquad z\approx \infty\,.
\ee 
Therefore the monodromy matrix expands as
\be\label{eq:Omega-inf}
\Omega(z,\tau)= 1+\frac{Q^R}{z}+\mathcal O(z^{-2}),\qquad\qquad z\approx \infty
\ee
where we have defined
\be\label{eq:qr0} 
Q^R=\int_0^{2\pi}d\sigma\ \tilde J_\tau(\sigma).
\ee
Here $Q^R$ is the (algebra-valued) conserved charge associated to the invariance of the PCM action under right-multiplication ($\tilde g\to \tilde gh_R^{-1}$ in~\eqref{eq:globPCM}). 
 As expected, $Q^R$ coincides with the expression obtained by applying Noether's theorem to~\eqref{eq:SYB} and, remarkably, it is conserved also in the twisted model because $\tilde J$ is periodic. In other words, {around $z= \infty$} the situation is completely analogous to the familiar case of the periodic PCM. In fact, right-multiplication remains a symmetry of the hYB model ($g\to gu_R^{-1}$ in~\eqref{eq:globYB}).

New features, instead, appear when expanding around $z=0$. 
It is convenient to first do a gauge transformation of the Lax connection as 
\be\label{eq:Lp}
\mathcal L'= \tilde g \mathcal L\tilde g^{-1} -d\tilde g\tilde g^{-1},
\ee so that
$\mathcal L_\sigma'= \frac{1}{1-z^2} \left(z\partial_\tau\tilde g\tilde g^{-1}+z^2\partial_\sigma\tilde g\tilde g^{-1}\right)$.
In terms of~$\mathcal L'$, the monodromy matrix can be written as 
\be
\Omega(z,\tau) = \tilde g(2\pi)^{-1}\mathcal P\exp\left(-\int_0^{2\pi}d\sigma'\mathcal L_\sigma'(z,\tau,\sigma')\right)\tilde g(0).
\ee
In the periodic case, the above is a similarity transformation that therefore does not change the eigenvalues, and the $\tilde g(2\pi)^{-1}$ and $\tilde g(0)$ can be ignored. However, now $\tilde g$ is not periodic.  It is useful to define a \emph{modified} monodromy matrix
\be
\Omega'(z,\tau)\equiv\tilde g(0)\Omega(z,\tau)\tilde g(0)^{-1} =W^{-1}\ \mathcal P\exp\left(-\int_0^{2\pi}d\sigma'\mathcal L_\sigma'(z,\tau,\sigma')\right),
\ee
where we used $W^{-1}=\tilde g(0) \tilde g(2\pi)^{-1}$.
The eigenvalues of $\Omega'(z,\tau)$ and of $\Omega(z,\tau)$ are the same because they are related by a similarity transformation, and they provide the usual infinite tower of conserved quantities. To study the asymptotics of the monodromy matrix around $z=0$, it is actually more convenient to work with $\Omega'(z,\tau)$, because the information will be given in terms of the twist $W$ rather than $\tilde g(2\pi)^{-1}$ and $\tilde g(0)$ separately.
In particular, since $\mathcal L'$ vanishes at $z=0$, one immediately finds
\be
\Omega'(z=0,\tau)=W^{-1},
\ee
which generalises the well-known periodic case result, where it reduces to the identity. Therefore, the (modified) monodromy matrix evaluated at $z=0$ encodes the information on the twisted boundary conditions controlled by $W$. Expanding to the next order around $z=0$ one finds
\be\label{eq:Omega-0}
\Omega'(z,\tau)=W^{-1}-z\ W^{-1}\check Q^L+\mathcal O(z^2),\qquad\qquad z\approx 0,
\ee
where we have defined
\be\label{eq:ql0} 
\check Q^L=\int_0^{2\pi}d\sigma\ \partial_\tau\tilde g\tilde g^{-1}.
\ee
The check on top of $\check Q^L$  is meant to  remind that, in the presence of the twist, these quantities are generically \emph{not}  conserved,\footnote{While we still have $\partial_\alpha(\partial^\alpha\tilde g\tilde g^{-1})=0$, this does not imply that $\check Q^L$ is conserved because $\bar{\tilde J}_\alpha\equiv\partial_\alpha\tilde g\tilde g^{-1}=\tilde g\tilde J_\alpha\tilde g^{-1}$ satisfies twisted boundary conditions, so that $\partial_\tau \check Q^L=\bar{\tilde J}_\sigma(2\pi)-\bar{\tilde J}_\sigma(0)\neq 0$. It is actually not difficult to write down a non-local expression like 
\be \label{eq:deftQL}
\tilde Q^L = \int_0^{2\pi}d\sigma\ W^{\frac{2\pi-\sigma}{2\pi}}\left(\bar{\tilde J}_\tau(\tau,\sigma)-\frac{1}{2\pi}\left[\log W,\int^\tau_{\tau_0} d\tau'\bar{\tilde J}_\sigma(\tau',\sigma)\right]\right)W^{\frac{\sigma-2\pi}{2\pi}},
\ee
that is conserved ($\partial_\tau \tilde Q^L =0$) and that reduces to the local expression for the Noether charges of the periodic PCM ($\tilde Q^L = \int_0^{2\pi}d\sigma\ \bar{\tilde J}_\tau(\tau,\sigma)$) in the untwisted/undeformed limit $\eta\to 0$.
} and they may correspond to conserved changes only when considering the periodic case or, equivalently, the undeformed limit $\eta\to 0$. 

Let us finally make the following side, although important, remark that will be useful later. If we denote by $T_{\bar a}$ the generators of the Lie algebra of the group of left symmetries of the hYB model (cf.~\eqref{eq:globYB} and~\eqref{eq:condYB}) so that 
\be\label{eq:compR}
\Ad_{T_{\bar a}}R=R\Ad_{T_{\bar a}},
\ee
then the corresponding Noether charges  are given by $Q^L=Q^L_{\bar a}T^{\bar a}$ with
\be\label{eq:Q-Noether}
Q^L_{\bar a}=  \int_0^{2\pi}d\sigma\ \Tr[T_{\bar a}\ gA_\tau g^{-1}].
\ee
While  these are local expressions in terms of the field $g$ of the hYB model, they will in general be non-local when expressed in terms of $\tilde g$, the twisted PCM field.

The conclusion is that the monodromy matrix of the hYB deformation remains  analytic at $z=0$ and $z=\infty$, as expected. However, we anticipate that this does not ensure that the quasimomenta are also analytic there, as they may have branch points at $z=0$ (while  at $z=\infty$ this is not possible because there the expansion of the  monodromy  starts with $1$).

\subsection{Deformed/twisted (super)coset models} \label{sec:sup-cos}
The aim of this section is to show that the results that we found in the case of deformations/twists of the  PCM can be straightforwardly generalised  to the case of (semi)symmetric space sigma-models. Readers who are not interested in (super)cosets may skip this section as it is not needed for section~\ref{sec:ex}.

We will first discuss the case of (bosonic) cosets. Our starting point is a real Lie group $G$ and a subgroup $H\subset G$ so that the algebra $\alg g = \alg g^{(0)}+\alg g^{(1)}$ admits the $\mathbb Z_2$ grading where $[\alg g^{(i)},\alg g^{(j)}]\subset \alg g^{(i+j  \text{ mod }  2)}$, and we are identifying the Lie algebra of $H$ as $\alg h=\alg g^{(0)}$.
In the rest, unless explicitly stated otherwise, we will use a notation such that for example $\tilde J^{(i)}\in \alg g^{(i)}$, and $A^{(i)}\in \alg g^{(i)}$.
The action of  the $\sigma$-model on the coset $G\setminus H$ may be written as\footnote{ To keep notation simple,   in this section we will not introduce explicitly a worldsheet metric. The formulas that we write are then valid when the conformal gauge is used. Alternatively, it is easy to translate these formulas to those before fixing any gauge. It is enough to replace vectors $V_\pm$ with $V_+\to P^{\alpha\beta}_{(+)}V_\beta$ and $V_-\to P_{(-)\alpha}{}^{\beta}V_\beta$, where $P_{(\pm)}^{\alpha\beta}=\tfrac12 (\gamma^{\alpha\beta}\pm\epsilon^{\alpha\beta})$ are projectors, and worldsheet indices are raised and lowered with the combination $\gamma^{\alpha\beta}=\sqrt{|h|}h^{\alpha\beta}$ of the worldsheet metric $h_{\alpha\beta}$. See e.g.~\cite{Arutyunov:2009ga,Delduc:2013qra,Borsato:2017qsx}. Obviously, one may reinstate a worldsheet metric also in the case of PCM, but we only  point out this possibility here because of the applications of (super)cosets in string theory.} 
\be\label{eq:S-coset}
S = -T\int d^2\sigma \ \Tr\left(\tilde J_+ \hat d\tilde J_-\right),
\ee
where the currents are still given by $\tilde J_\pm=\tilde g^{-1}\partial_\pm \tilde g$ in terms of $\tilde g\in G$, and $\hat d=P^{(1)}$ projects on the subspace $\alg g^{(1)}=\alg g\setminus \alg h$. The above is in fact a minimal modification of the PCM action, in which the projector $\hat d$ is inserted. Because of this, the action has in fact a local invariance under right multiplication $\tilde g\to \tilde gh$ with $h\in H$, which allows to take $\tilde g\in G\setminus H$ in the coset. We will however never fix a gauge for this invariance, and prefer to think of $g$ as a generic element of $G$. The classical integrability is ensured by the Lax connection $\mathcal L_\pm =\tilde J^{(0)}_\pm + z^{\mp 1} \tilde J^{(1)}_\pm$.
Taking $g\in G$, the hYB deformation of the above action is \cite{Delduc:2013fga,Delduc:2013qra}
\be\label{eq:SYB-coset}
S_{\text{YB}}= -T \int d^2\sigma\  \Tr\left(J_+\ \hat d\frac{1}{1-\eta R_g\hat d}\ J_-\right),
\ee
which is still invariant under local $H$-valued right-multiplications.
Integrability is preserved in the deformed model and the Lax connection is now given by  $\mathcal L_\pm =  A^{(0)}_\pm + z^{\mp 1}  A^{(1)}_\pm$ where
\be\label{eq:Apm-coset}
 A_\pm =\frac{1}{1\pm\eta R_g\hat d}\  g^{-1}\partial_\pm g.
\ee 
Compared to the corresponding expressions~\eqref{eq:SYB} and~\eqref{eq:Apm} valid for the hYB of PCM, we see that the only modification comes from the insertion of the projector $\hat d$ in proper places.

{The case of the supercoset is formally identical. We now consider a real Lie supergroup $G$ whose algebra  admits the $\mathbb Z_4$ grading $\alg g = \alg g^{(0)}+\alg g^{(1)}+\alg g^{(2)}+\alg g^{(3)}$ so that $[[\alg g^{(i)},\alg g^{(j)}]]\subset \alg g^{(i+j \text{ mod } 4)}$, where $[[ \ , \ ]]$ is the superbracket (that coincides with the anti-commutator when both $\alg g^{(i)},\alg g^{(j)}$ have odd grading, and  the commutator otherwise). Notice that if $x,y$ belong to the Grassmann enveloping algebra of $\alg g$, then the bracket that must be used  is always given by the standard commutator $[x,y]$, because the Grassmann variables take care of the additional minus signs.
If we identify $H$ with the Lie group of $\alg h=\alg g^{(0)}$, then the action of the $\sigma$-model on the supercoset $G\setminus H$ is still given by~\eqref{eq:S-coset}, with the exception that now $\hat d=\tfrac12 P^{(1)}+P^{(2)}-\tfrac12 P^{(3)}$ and that the trace should be substituted by the supertrace~\cite{Arutyunov:2009ga}. With these minor substitutions, the hYB deformation of the supercoset action is  still given by~\eqref{eq:SYB-coset} \cite{Delduc:2013qra}, and the Lax connection is the obvious generalisation of the supercoset Lax connection $\mathcal L_\pm =  A^{(0)}_\pm +z A^{(1)}_\pm+ z^{\mp 2}  A^{(2)}_\pm+z^{-1} A^{(3)}_\pm$, where $A_\pm$ is given again by~\eqref{eq:Apm-coset}. In the following, we will be able to treat the coset and the supercoset cases at the same time, by exploiting the fact that the only difference is in the definition of the linear operator $\hat d$, and that we can always work with the standard commutator if we interpret all formulas as being valid for the Grassmann enveloping algebra of $\alg g$. 
}

\subsubsection{The twist}
Because the Lax connections of the (super)coset and of its hYB deformation take again the same form, one may  study an on-shell identification induced by demanding
$\tilde J_\pm=A_\pm$, where $\tilde J=\tilde g^{-1}d\tilde g$ is the Maurer-Cartan form of the twisted (super)coset with field $\tilde g\in G$, and $A_\pm$ is now given by~\eqref{eq:Apm-coset}. 
As done before, we can still relate $g$ and $\tilde g$ by the twist field $\mathcal F$ as $g=\mathcal F \tilde g$.\footnote{Let us point out that the identifications $g=\mathcal F \tilde g$ and $\tilde J_\pm=A_\pm$ will only be valid in a specific gauge, which may be different for the coset representatives  ${g}$ and $\tilde{g}$. In general we must write their relation as $g=\mathcal F \tilde g h$ for some $h\in H$ and then the on-shell identification will read $h^{-1} \tilde{J}_\pm h + h^{-1} \partial_\pm h = A_\pm$.} By repeating precisely the same steps as in the PCM case, with minor modifications given by the presence of $\hat d$, we  therefore find that the twist field must satisfy\footnote{In terms of the deformed variables, one  finds the relation
\begin{equation} 
\partial_\pm {\cal F} = V_\pm {\cal F} , \qquad V_\pm = \pm \eta R \text{Ad}_g \hat{d} A_\pm \ .
\end{equation}
}
\be\label{eq:dF3-coset}
\eta^{-1} \omega_{\mathcal{F}}(\mathcal F^{-1} d\mathcal F)=P^T\ \hat d_{{\tilde g}^{-1}}(*d\tilde g\tilde g^{-1}),
\ee
where we defined
\be
\hat d_{{\tilde g}^{-1}} \equiv \AD_{\tilde g}\hat d\AD_{\tilde g}^{-1}.
\ee
This is once again a minor modification of the equation~\eqref{eq:dF3} found for the PCM case. Notice that the left-hand-side of~\eqref{eq:dF3-coset} is the same as that of~\eqref{eq:dF3}, so we can proceed  to solve the equation in the same way. We write $\mathcal F=\exp(RX)$ and define $Y\in \alg f^*$ as in~\eqref{eq:red-y-F} and~\eqref{eq:red-y-X}. We will therefore still have equation~\eqref{eq:dydf}, which allows us to write
\be\label{eq:y-tg-coset}
dY=P^T\ \hat d_{{\tilde g}^{-1}} (*d\tilde g\tilde g^{-1}).
\ee 
This equation is now solved by
\be\label{eq:soly-coset}
Y(\tau,\sigma) = P^T\left(\int_0^\sigma d\sigma' \hat d_{{\tilde g}^{-1}}(\partial_\tau\tilde g\tilde g^{-1}) (\tau,\sigma')+\int_0^\tau d\tau' \hat d_{{\tilde g}^{-1}}(\partial_\sigma\tilde g\tilde g^{-1} )(\tau',0)\right)+Y(0,0).
\ee
As before, $\tilde g$ satisfies twisted boundary conditions
\be
\tilde g(2\pi)=W\tilde g(0),\qquad W=\mathcal F(2\pi)^{-1}\mathcal F(0).
\ee
To summarise, the twist in the (super)coset case has a solution that is very similar to that of  the PCM case. The only difference comes in the solution for $Y$, with an insertion of $\hat d_{{\tilde g}^{-1}}$ as in~\eqref{eq:soly-coset}.
{Let us remark that in the supercoset case the solution for the twist is valid also for subalgebras $\alg f$ with non-trivial subspaces of odd grading, because the calculations have been carried out in the Grassmann enveloping algebra of $\alg g$.}

\subsubsection{The monodromy matrix}
The Lax connection of the (super)coset has simple (and double) poles at $z=0$ and $z=\infty$, that induce  essential singularities of the monodromy matrix.  Around these singular points, one may diagonalise the monodromy matrix using a generalisation of the method recalled in footnote~\ref{f:Omdiag} that takes into account the underlying algebraic (semi)symmetric space structure (see for example~\cite{Hollowood:2010dt,Hollowood:2011fq,Hollowood:2019ajq}). In the case of the undeformed and periodic (super)coset, the expansion around $z=1$ (or equivalently $z=-1$) yields the global charges corresponding to left $G$-valued transformations (which in the PCM case were found at $z=0$)~\cite{Arutyunov:2009ga}. We will now show that it is still possible to expand the monodromy matrix around $z=1$ also in the deformed/twisted (super)coset. To avoid repeating the discussion twice, we do this directly for the supercoset, the corresponding results for the coset case are easily obtained by setting the components of odd grading to zero, and by doing a simple redefinition of the spectral parameter ($z^2\to z$).

The monodromy matrix  $\Omega(z,\tau) = \mathcal P\exp(-\int_0^{2\pi}d\sigma\ \mathcal L_\sigma)$ is now written in terms of the Lax connection
\be
\mathcal L_\sigma = \tilde J^{(0)}_\sigma +z\, \tilde J^{(1)}_\sigma+\frac12 \left(\frac{1}{z^2}-z^2\right)\tilde J^{(2)}_\tau+\frac12 \left(\frac{1}{z^2}+z^2\right)\tilde J^{(2)}_\sigma+\frac{1}{z}\, \tilde J^{(3)}_\sigma.
\ee
After applying a gauge transformation $\mathcal L'_\sigma= \tilde g\mathcal L_\sigma \tilde g^{-1} - \partial_\sigma \tilde g \tilde g^{-1}$ to  the Lax connection one finds
\be
\mathcal L'_\sigma = (z-1)\, a^{(1)}_\sigma+\frac12 \left(\frac{1}{z^2}-z^2\right) a^{(2)}_\tau+\frac12 \left(\frac{1}{z^2}+z^2-2\right) a^{(2)}_\sigma+\left(\frac{1}{z}-1\right)\, a^{(3)}_\sigma,
\ee
where $a^{(i)}= \tilde g\tilde J^{(i)} \tilde g^{-1}= P^{(i)}_{{\tilde g}^{-1}} (d\tilde g \tilde g^{-1})$. Notice that, despite the notation, $a^{(i)}$ does not necessarily belong to $\alg g^{(i)}$ only, but in general to the whole $\alg g$.
Under such gauge transformation, the monodromy is rewritten as before as $\Omega(z) = \tilde g(2\pi)^{-1}\mathcal P\exp\left(-\int_0^{2\pi} {d\sigma} \ \mathcal L_\sigma'\right)\tilde g(0)$, and we can define again
\be
\Omega'(z)=\tilde g(0)\Omega(z)\tilde g(0)^{-1} =W^{-1}\mathcal P\exp\left(-\int_0^{2\pi}{d\sigma} \ \mathcal L_\sigma'\right).
\ee
If we write $z=1+\epsilon$ we see that $\mathcal L_\sigma' = \epsilon (a^{(1)}_\sigma-2a^{(2)}_\tau-a^{(3)}_\sigma) +\mathcal O(\epsilon^2)$,
so that in the presence of the deformation/twist
\be
\Omega'(z=1+\epsilon) =W^{-1}+\epsilon \ W^{-1}\int_0^{2\pi} {d\sigma} \ (-a^{(1)}_\sigma+2a^{(2)}_\tau+a^{(3)}_\sigma) +\mathcal O(\epsilon^2).
\ee
Similar comments valid in the PCM case now apply also for the (super)coset. When $W=1$, the integral at $\mathcal O(z)$ can be identified with the Noether charge for the global $\alg g$ isometry. When $W\neq 1$, instead, while there  still exists a divergenceless current as a consequence of the flatness of the Lax connection, this does not generically give rise to conserved quantities because the current does not satisfy periodic boundary conditions.

\section{Twists and classical spectral curves}\label{sec:ex}
In this section we will explicitly construct the twist  for various  examples of hYB deformations of the PCM, namely abelian, almost-abelian and Jordanian  deformations. In the abelian and Jordanian case, we show how the finite-gap equations and {their so-called ``one-cut solutions''} change when turning on the deformation parameter $\eta$, i.e.~for the twisted models. This latter discussion closely follows that of appendix \ref{app:CSC} 
{where we give  in some detail the construction of the one-cut solutions for the resolvents of the undeformed and periodic PCM models based on the $U(2)$ and $SL(2,\mathbb{R})$ groups.} As mentioned in the appendix, we may want to reinstate a worldsheet metric, and in that case we would also need to impose the Virasoro constraints. These may be seen as some extra conditions that fix one of the free parameters of our solutions. This observation will suffice for our discussion, and we will not discuss  the Virasoro constraints further. 

Before diving into the explicit examples and the construction of the finite-gap equations, let us give a general remark which distinguishes the deformed{/twisted} case from the undeformed{/periodic} one.
As  discussed in section \ref{sec:yb-int}, to describe the CSC we need the information of the quasimomenta at its asymptotic values, $z=\infty$ and $z=0$. This  requires {the knowledge of  the expansions of the eigenvalues of}  the  monodromy matrices given in \eqref{eq:Omega-inf} and \eqref{eq:Omega-0}. For the monodromy around infinity \eqref{eq:Omega-inf}, which involves the Noether charge $Q_R$, we can always use the (unbroken) global $G_R$ invariance of the twisted model to diagonalise it. For the monodromy around $z=0$ \eqref{eq:Omega-0} this is, however, not possible  in general because the global $G_L$ symmetries are broken. {We will have to start by studying the eigenvalues of the twist $W^{-1}$, and then their $\mathcal O(z)$ correction induced by the term $W^{-1} \check{Q}_L$, where $\check{Q}_L$ is in general not conserved. When the quasimomenta have asymptotics as in~\eqref{eq:as-p}, we can use the results of~\cite{Dorey:2006zj,Vicedo:2008ryn,Krichever:1996ut} reviewed in section \ref{sec:yb-int}, and conclude that this data is enough to reconstruct the quasimomenta everywhere on the Riemann surface.} As mentioned, we will illustrate this fact by reconstructing the quasimomenta/resolvent in the case of one branch cut.

\subsection{Abelian}\label{sec:ab}
An $R : \mathfrak{f}^\star \rightarrow \mathfrak{f}$ is said to be abelian if  $\mathfrak{f}$ is an abelian algebra, see e.g.~\cite{vanTongeren:2015soa}. We call $R$ abelian and diagonal if the basis of $\alg g$ can be chosen such that $\mathfrak{f}$ is furthermore spanned only by Cartan generators of $\mathfrak{g}$, that can be represented by diagonal matrices. $R$ is abelian and non-diagonal when this is not possible, which implies that $W$ is not diagonalisable as a matrix and it can only  be put in Jordan block form. After constructing the twist for the general abelian case, we will treat the CSC discussion of these two cases separately. \\

\noindent \textbf{The twist $W$} --- For simplicity, let us consider here the case of a rank-2 $R$-matrix, higher ranks being a straightforward generalisation. In other words,
\begin{equation}
{r} =  T_1 \wedge T_2 , \qquad [T_1 , T_2] = 0 \ ,
\end{equation}
where $T_i, i=1,2$ are the generators of $\mathfrak{f}$. Taking $X,Y \in \mathfrak{f}^\star$  we have from \eqref{eq:red-y-X}  simply that $X = \eta Y${, because when $\alg f$ is abelian we can use the identity  $P^T[\alg f,\alg g]=T^i\Tr(T_i[\alg f,\alg g])=T^i\Tr([T_i, \alg f]\alg g])=0$. Then ${\cal F} = \exp (\eta R Y )$ so that the twist becomes
\begin{equation}
W = \exp \left(-\eta R {\cal Q} \right) , \qquad {\cal Q} = Y( 2\pi ) - Y ( 0) = { \int^{2\pi}_0 d\sigma \ P^T(\partial_\tau \tilde{g}\tilde{g}^{-1} )}\ .
\end{equation}
Because of the abelian nature of $\mathfrak{f}$, it immediately follows that $T_i$ satisfies the compatibility condition of \eqref{eq:compR} with $R$, such that the Noether charges $Q_i^L$ defined as in~\eqref{eq:Q-Noether} are conserved.\footnote{In addition, the abelian twist satisfies $W T_i W^{-1} = T_i$ and one can show that this implies that $\tilde{Q}_i^L$, with $\tilde{Q}^L$ defined in \eqref{eq:deftQL}, is a \textit{local} conserved quantity which in fact coincides with $Q_i^L$.} In fact, they coincide with the projections $\check{Q}_i^L=\Tr(T_i\check{Q}^L)$ as follows from
\begin{equation} \label{eq:abLcharges}
Q_i^L =  \int^{2\pi}_0 d\sigma \Tr \left[ T_i \ g  A_\tau g^{-1} \right ] =  \int^{2\pi}_0 d\sigma \Tr \left[ T_i \ \tilde{g} {\tilde J}_\tau \tilde{g}^{-1} \right ]  = \check{Q}_i^L,
\end{equation}
when writing $g=\mathcal F\tilde g$, and thus we may write 
\begin{equation} \label{eq:abcalQ}
{\cal Q} =   Q_i^L T^i =  \check{Q}_i^L T^i \ .
\end{equation}
 A different but equivalent discussion of the solution for the twist in the abelian case can be found in \cite{vanTongeren:2018vpb}. \\

\subsubsection{Abelian diagonal deformation} \label{ss:exabdiag}

 To look at diagonal abelian deformations and to be as explicit as possible, it is convenient to consider the compact $U(2)$ group, which admits only abelian deformations \cite{Lichnerowicz1988}. One may compare our discussion to that of~\cite{Frolov:2005ty}, where a TsT transformation of $S^3$ was considered and the corresponding finite-gap equations were written. However, let us remark that there the TsT transformation involved the Cartans of each of the two copies $SU(2)_L\times SU(2)_R$ of the PCM, which is different from our setup given that we only deform the left $G$ symmetry of the PCM. For the CSC discussion of the undeformed (periodic) PCM based on $U(2)$ and our conventions on its generators we refer to appendix \ref{app:CSC}. Here we turn on the deformation/twist and take\footnote{By a similarity transformation the generators in the definition of $R$ can always be taken in the Cartan subalgebra.}
 \begin{equation}
{r} = T_0 \wedge T_3 \ ,
 \end{equation}
so that the deformation  preserves the Cartan isometries.
 Explicitly, the twist then is {
 \begin{equation}
 W = \begin{pmatrix}
  e^{-\frac{i \eta}{2} (Q^L_3 - Q^L_0 )} & 0 \\
 0 &  e^{ -\frac{i \eta}{2}(Q^L_3 + Q^L_0 )}  
 \end{pmatrix} \ .
 \end{equation}
}%

\noindent \textbf{Asymptotics of quasimomenta} --- Since the expansion of the quasimomenta at infinity does not change compared to the undeformed case, it will behave as in the appendix, eq.~\eqref{eq:U2QMInf}. At $z =0$, however, the expansion in the case of twisted models introduces the quantity $\check{Q}^L$ which  we cannot assume  of diagonal form. For $\mathfrak{u}(2)$ it is
{
\begin{equation}
\check{Q}^L=
-i \left(
\begin{array}{cc}
\check{Q}^L_0+\check{Q}^L_3 & \check{Q}^L_1-i \check{Q}^L_2\\
\check{Q}^L_1+i \check{Q}^L_2 & \check{Q}^L_0-\check{Q}^L_3
\end{array}
\right).
\end{equation}
}%
Since we are after the eigenvalues of the monodromy matrix, we will have to apply a similarity transformation $S$ to the full $\Omega' $ to diagonalise it. Assuming $S = S^{(1)} + z \, {S^{(2)}} + {\cal O}(z^2)$ then $S^{(1)}$ diagonalises $\Omega' (z=0) = W^{-1}$, and at order $z$ we get the condition
\begin{equation}
{
(S^{(1)})^{-1} W^{-1}S^{(2)}-(S^{(1)})^{-1} S^{(2)}(S^{(1)})^{-1} W^{-1} S^{(1)}-  (S^{(1)})^{-1} W^{-1} \check{Q}^L S^{(1)} = D ,}
\end{equation}
where $D$ is a diagonal matrix. In this case, $W$ is already diagonal and we can take $S^{(1)} = \mathbf{1}$, so that the condition reduces to
\begin{equation}
[W^{-1} ,S^{(2)}]-W^{-1} \check{Q}^L = D.
\end{equation}
With $W$ diagonal, the diagonal part of $S^{(2)}$ will not participate in this equation. The off-diagonal part of $S^{(2)}$ is taken such that  the off-diagonal contribution of $W^{-1} \check{Q}^L$ is canceled by the commutator. Therefore, after diagonalisation we have
\begin{equation}
D =  i  \left(
\begin{array}{cc}
e^{ \frac{i \eta}{2} (Q^L_3 - Q^L_0 )} \left({Q}^L_0+{Q}^L_3\right) & 0\\
0 & e^{ \frac{i \eta}{2} (Q^L_3 + Q^L_0 )}  \left({Q}^L_0-{Q}^L_3 \right)
\end{array}
\right).
\end{equation}
Note that, because of \eqref{eq:abLcharges}, we can drop the checks everywhere. The expansion of the quasimomenta around $z=0$ is thus
\begin{equation}
\begin{aligned}
p_1(z) &\approx 2\pi m_1 -\frac{\eta}{2} (Q^{L}_0-Q^{L}_3)+ z (Q^{L}_0+Q^{L}_3),\\
p_2(z) &\approx 2\pi m_2 +\frac{\eta}{2} (Q^{L}_0+Q^{L}_3)+ z (Q^{L}_0-Q^{L}_3).
\end{aligned}
\end{equation}
{In the undeformed and periodic case, the fact that only the components $Q^{L}_0,Q^{L}_3$ (and not $Q^{L}_1,Q^{L}_2$) appear at $\mathcal O(z)$ is a consequence of the presence of the unbroken $G_L$ isometry, that allows one to diagonalise the Noether charge. While $G_L$ is broken after the deformation/twist, the same components $Q^{L}_0,Q^{L}_3$ are singled out at $\mathcal O(z)$, but now because of the particular form of the twist $W$ chosen.}
According to the discussion given in section \ref{sec:yb-int}, one can now reconstruct the quasimomenta and resolvents on the full Riemann surface, provided that the number and location of branch cuts is specified. \\

\noindent \textbf{One-cut resolvents} --- Let us illustrate the reconstruction of the resolvent for the $U(2)$ twisted model in the case of a single branch cut of finite size. This discussion closely follows the one for the untwisted (undeformed) model of appendix \ref{app:CSC}. In fact, we take an identical ansatz for $G_1(z)$ and $G_2(z)$ given in \eqref{eq:U2G1}--\eqref{eq:u2constraint}. This implies that the condition for $G_A(z)$ to be analytical everywhere leads to the same solutions for $q_A^{\pm}$ given in \eqref{eq:U2qpm} and $B$ and $C$ given in \eqref{eq:U2SolBC} in terms of the parameters $\epsilon_i, i=\{1, \ldots , 4\}$, $\alpha_\pm$ and $A$. Since the asymptotic behaviour at infinity is identical to the undeformed case as well, we will also have the same solution for $A$, given in \eqref{eq:U2SolA}, and for $Q_0^R$ and $Q_3^R$ given in \eqref{eq:U2SolQR}. Around $z=0$, however, the deformation will start to play a role.  For $\eta\neq 0$, it will be simplest to solve the equations that match the leading term around $z=0$ in terms of $Q_0^L$ and $Q_0^R$ (in the undeformed  $\eta = 0$ case this would not be possible, as $Q_0^L$ and $Q_0^R$ do not appear at this order). At this stage we find
\begin{equation} \label{eq:u2defQL}
\begin{aligned}
Q_0^L &= \frac{1}{\eta} \left( 4 \sqrt{C} \left( \frac{\alpha_+}{\epsilon_1 + \epsilon_3} + \frac{\alpha_-}{\epsilon_2 + \epsilon_4} \right)  + 2(m_1 - m_2 - n) \pi  \right) ,  \\
 Q_3^L &= \frac{2}{\eta} \left( \alpha_+ + \alpha_- -  (m_1 + m_2) \pi \right)  ,
\end{aligned}
\end{equation}
where $C = C (\epsilon_i, \alpha_\pm , n)$ may be substituted by the solutions \eqref{eq:U2SolBC} and \eqref{eq:U2SolA}.  Note that we should make sure that the undeformed limit can be taken appropriately, which we will do momentarily. The equations matching the subleading term at $z=0$ now become, after substituting the solutions \eqref{eq:U2SolBC}, \eqref{eq:U2SolA} and \eqref{eq:u2defQL}, constraint equations between the remaining continuous variables $\alpha_\pm$ and $\epsilon_i, i=\{1, \ldots , 4\}$. Writing them for simplicity in terms of $A = A(\epsilon_i, \alpha_\pm , n), B=B(\epsilon_i), C = C(\epsilon_i, \alpha_\pm , n)$, and substituting the solutions \eqref{eq:u2defQL}, they are found to be
\begin{equation} \label{eq:u2defconstraints}
\begin{aligned}
   2(m_1 -  m_2- n)\pi   - \eta ( \alpha_+ - \alpha_- ) + 4 \sqrt{C} \left( \frac{\alpha_+}{\epsilon_1 + \epsilon_3} + \frac{\alpha_-}{\epsilon_2 + \epsilon_4} \right)&=0 , \\
 2(\alpha_+ + \alpha_-) -  2(m_1+m_2  ) \pi  +\frac{\eta(B-2 C ) \alpha_-}{\sqrt{C} (\epsilon_2 + \epsilon_4)} + \frac{\eta(B+ 2 C ) \alpha_+}{\sqrt{C} (\epsilon_1 + \epsilon_3)} &=0 , 
 \end{aligned}
\end{equation}
The first constraint equation implies that we can simplify the expression for $Q_0^L$ as $Q_0^L = \alpha_+ - \alpha_-$ and so it coincides precisely with $Q_0^R$, as it should (see also  the comments given at the end of case $(i)$ of appendix \ref{app:CSC}).
As in the undeformed case, we therefore effectively have three remaining continuous variables within our class of solutions, possibly further constrained by the Virasoro constraints. {To finally obtain the full quasimomenta and resolvents, we must simply substitute the solutions for $q_A^\pm, A, B$ and $C$ in \eqref{eq:U2G1} and \eqref{eq:U2G2},  and subject them to the constraints \eqref{eq:u2defconstraints}.}

Let us briefly discuss the undeformed limit $\eta \rightarrow 0$. The constraints \eqref{eq:u2defconstraints}  simply reduce to
\begin{equation} \label{eq:u2defconstraintslimit}
4 \sqrt{C} \left( \frac{\alpha_+}{\epsilon_1 + \epsilon_3} + \frac{\alpha_-}{\epsilon_2 + \epsilon_4} \right)  + 2(m_1 - m_2 - n) \pi =0 , \qquad \alpha_+ + \alpha_- -  (m_1 + m_2) \pi  = 0 \ ,
\end{equation}
and thus in the undeformed limit both the numerators and denominators of $Q^L_i $ in  \eqref{eq:u2defQL} tend to zero. In addition, these equations imply precisely the same expression  for $\alpha_+$ given in \eqref{eq:U2Solap} and for  $\sqrt{C}$ given in \eqref{eq:U2SolsC}, where the latter implied the quartic constraint equation \eqref{eq:U2CQ4} of the undeformed case. To obtain a finite expression for $Q_i^L$, we can simply use \eqref{eq:u2defconstraints}  in \eqref{eq:u2defQL}, and then we will obtain precisely the same (undeformed) expressions for $Q_i^L$ as given in \eqref{eq:U2SolQR}.

\subsubsection{Abelian non-diagonal} \label{ss:abnondiag}
Let us now look at a non-diagonal abelian deformation. We will have to take a non-compact algebra, and to be as explicit as possible, while remaining simple, we choose to work with $\mathfrak{gl}(2, \mathbb{R})$. As for our conventions, we take the same generators as those for $\mathfrak{sl}(2,\mathbb{R})$ given in \eqref{eq:GenSL2} and we add another compact direction $T_0 = \frac{1}{2}\mathbf{1}$ (with $T^0 = 2 T_0$).   We  then choose
\begin{equation}
r = T_+ \wedge T_0 ,
\end{equation}
so that $\mathfrak{f} = \text{span}\{T_+ , T_0\}$ and $\mathfrak{f}^\star = \text{span}\{ T^+ , T^0 \}$. Then ${\cal Q} =  \left( Q_+^L T^+ + Q_0^L T^0 \right)$ and the twist becomes
\begin{equation}\label{eq:Wnda}
{
W= e^{\frac{ \eta }{2}Q^L_+}\left(
\begin{array}{cc}
1 & -\eta Q^L_0   \\
 0 & 1\\
\end{array}
\right).}
\end{equation}

\noindent {\bf Asymptotics of the quasimomenta} --- Around $z=\infty$ the quasimomenta behave  as 
\begin{equation}\label{eq:p-non-comp}
p_1(z) \approx - \frac{i}{z} \left( Q^R_0 + Q^R_3 \right) , \qquad p_2(z) \approx  -\frac{i}{z} \left( Q^R_0 - Q^R_3 \right) , \qquad z \approx \infty \ .
\end{equation}
For the expansion around $z=0$ note that
 the twist is of Jordan block form and it is not diagonalisable. The degeneracy of the eigenvalues is  lifted when including  $z$-corrections, but the expansion of the quasimomenta around $z=0$ turns out to be (see appendix~\ref{app:JordanForm})
\begin{equation}\label{eq:p-as-non-diag}
\begin{aligned}
p_1(z)\approx 2\pi  m_1 +\frac{i\eta}{2}Q^L_+  +i \sqrt{z}\sqrt{-\eta Q^L_0Q^L_+} +i z Q^L_0,\\
p_2(z)\approx 2\pi  m_2 +\frac{i\eta}{2}Q^L_+  -i \sqrt{z}\sqrt{-\eta Q^L_0Q^L_+} +i z Q^L_0.
\end{aligned}
\end{equation}
The above expansion seems to be problematic because of the appearance of the $\sqrt{z}$ term. This does not fall within our assumptions (c.f.~\eqref{eq:as-p}) and, in particular, we cannot directly apply the arguments of section \ref{sec:yb-int} to claim that the data from $z=0$ and $z=\infty$ is enough to reconstruct the quasimomenta everywhere on the Riemann surface. In fact, $z=0$ should be a puncture, because it is a pole for the abelian integral $Q$ of~\cite{Dorey:2006zj,Vicedo:2008ryn,Krichever:1996ut}, but the above expansion shows that it should also be  a branch point. {
One may try to regularise the problem by modifying the choice for $Q$, so that the position of the pole is moved away from $z=0$ when $\eta\neq 0$, but this would introduce additional technical difficulties because one would need to expand the monodromy matrix around a point where the path-ordered exponential cannot be simplified. Another possibility may be to regularise the twist itself in such a way that the degeneracy of the eigenvalues of $W$ is lifted (in other words, under such regularisation it would be the branch point that is moved away from $z=0$). However, we believe that this strategy would intrinsically change the problem, and that such regularisation would not be useful to understand the nature of the CSC in this case.  While we are not able to say whether the above data is enough to reconstruct the quasimomenta on the full complex plane in the most generic set-up, it may  still be enough for subclasses of finite-gap solutions, see e.g.~\cite{Ouyang:2017yko}.} {In fact in what follows, we will  show that for this abelian non-diagonal case one can construct a one-cut resolvent which has the correct asymptotic behaviour.} \\

\noindent {\bf One-cut resolvent} --- Consider the ansatz for the resolvents 
\begin{align}
G_1(z) &= \pi n + \left(\frac{\alpha_+ - q_1^+}{1 - z}  +  \frac{\alpha_- + q_1^-}{1 + z} \right) -   \left(\frac{ \epsilon_1^{-1} q_1^+}{1 - z}  -  \frac{\epsilon_2^{-1} q_1^-}{1 + z} \right) \sqrt{Az^2+Bz} , \label{eq:ndaG1}\\
G_2(z) &= -\pi n + \left(\frac{\alpha_+ - q_2^+}{1 - z}  +  \frac{\alpha_- + q_2^-}{1 + z} \right) +  \left(\frac{ \epsilon_3^{-1} q_2^+}{1 - z}  -  \frac{\epsilon_4^{-1} q_2^-}{1 + z} \right) \sqrt{Az^2+Bz} \label{eq:ndaG2} ,
\end{align}
together with the condition  \eqref{eq:u2constraint}.
Compared to~\eqref{eq:U2G1} and~\eqref{eq:U2G2}  we  have $C=0$ since in this example one of the branch points of the single cut is forced to be located at $z=0$. This will  then change the procedure of solving the finite-gap equations compared to the appendix. Importantly, we will have additional equations from matching now also the ${\cal O}(\sqrt{z})$ term around $z=0$, next to the ${\cal O}(z^0)$ and ${\cal O}(z)$ terms.

 From requiring analyticity of the resolvents  everywhere, and in particular at $z=\pm 1$, as well as requiring \eqref{eq:u2constraint}, we find for $q^\pm_A$ the solutions given in \eqref{eq:U2qpm} and\footnote{Note that for a cut of finite size we may not have $A=0$, and thus we cannot have $\epsilon_1 = \epsilon_3$ and $\epsilon_2 = \epsilon_4$ at the same time.}
\begin{equation} \label{eq:AndSolAB}
A = \frac{1}{8} \left( (\epsilon_1 - \epsilon_3)^2  + (\epsilon_2 - \epsilon_4)^2 \right), \qquad B = \frac{1}{8} (\epsilon_1 + \epsilon_2 - \epsilon_3 - \epsilon_4) ( \epsilon_1 - \epsilon_2 - \epsilon_3 + \epsilon_4 ) \ .
\end{equation}
Around infinity, the leading behaviour of the quasimomenta does not differ compared to the appendix {(except for overall factors of $i$ due to the fact that we are comparing to the $U(2)$ example there)}  but since we already solved for the parameter $A$ we will have to solve the corresponding equations differently. We find that, after substituting \eqref{eq:U2qpm}, the two equations for the two resolvents at leading order, i.e. ${\cal O}((1/z)^0)$ coincide and can be solved as
\begin{equation} \label{eq:AndSolap}
\alpha_+ = \frac{(2 \sqrt{A} \alpha_- - (\epsilon_2 + \epsilon_4) n \pi) (\epsilon_1+ \epsilon_3)  }{2 \sqrt{A} (\epsilon_2 + \epsilon_4)} \ , 
\end{equation}
where $A = A(\epsilon_i)$. The ${\cal O}(1/z)$ contribution can be solved for $Q_0^R$ and $Q_3^R$ where the expressions are  (cf.~\eqref{eq:U2SolQR})
\begin{equation} 
\begin{aligned}
Q_0^R &= i( \alpha_- - \alpha_+ ), \\
Q_3^R &= i\frac{2 A  \left( (\epsilon_2 + \epsilon_4) \alpha_+ + (\epsilon_1 + \epsilon_3 )\alpha_- \right) + B  \left( (\epsilon_2 + \epsilon_4) \alpha_+ - (\epsilon_1 + \epsilon_3 )\alpha_- \right)  }{(\epsilon_1 + \epsilon_3) (\epsilon_2 + \epsilon_4) \sqrt{A}}  ,  
\end{aligned}
\end{equation}
and  where it is understood that $\alpha_+ = \alpha_+ (\epsilon_i, n)$ as in \eqref{eq:AndSolap}, and $A=A(\epsilon_i)$ and $B=B(\epsilon_i)$ as in \eqref{eq:AndSolAB}. Let us now turn to the equations around $z=0$. The leading contributions for the two resolvents force us to set
\begin{equation} \label{eq:AndSoln}
n =  m_1 -  m_2 \ ,
\end{equation}
and the remaining equation can be solved for $Q_+^L$ as
\begin{equation}\label{eq:AndSolQpL}
\eta Q_+^L = 2 i (m_1 + m_2)\pi + i \frac{(\epsilon_1 + \epsilon_3) (m_1 - m_2)\pi}{\sqrt{A}} -  \frac{2 i(\epsilon_1 + \epsilon_2 + \epsilon_3 + \epsilon_4)\alpha_-}{\epsilon_2 + \epsilon_4} \ ,
\end{equation}
where the solutions \eqref{eq:U2qpm} and \eqref{eq:AndSolap} have been substituted and it should be understood that $A = A(\epsilon_i)$. Note that this becomes a constraint equation  when $\eta = 0$. Next we can solve for the ${\cal O}(z)$ contributions around $z=0$. The two equations for the resolvents are identical and give 
\begin{equation} \label{eq:AndSolQ0L}
Q_0^L = i ( \alpha_- - \alpha_+ )  \ ,
\end{equation}
where $\alpha_+ = \alpha_+ (\epsilon_i, m_1, m_2)$ upon \eqref{eq:AndSolap} and \eqref{eq:AndSoln}.
Note that again $Q_0^L = Q_0^R$ as it must be for consistency. As the final step one must  solve the ${\cal O}(\sqrt{z})$ contribution. Upon the solutions \eqref{eq:U2qpm}, they again become identical. Substituting also the solutions \eqref{eq:AndSolap},  \eqref{eq:AndSoln}, \eqref{eq:AndSolQpL}, and\eqref{eq:AndSolQ0L} we find a complicated constraint equation between the variables $\alpha_-, \epsilon_i, m_1$ and $m_2$ which we for simplicity write as
\begin{equation}
2\sqrt{B} \left( \frac{ \alpha_+}{\epsilon_1 + \epsilon_3} + \frac{\alpha_-}{\epsilon_2 + \epsilon_4} \right) = -i  \sqrt{-\eta Q_0^L Q_+^L} , 
\end{equation}
with $B = B(\epsilon_i)$, $\alpha_+ = \alpha_+ (\epsilon_i, m_1, m_2)$, $Q_0^L = Q_0^L (\alpha_- , \epsilon_i , m_1, m_2)$ and $Q_+^L = Q_+^L (\epsilon_i , m_1, m_2)$.
We may think of this equation as fixing $\alpha_-$.  To conclude, we have constructed a one-cut resolvent with a consistent asymptotic behaviour that may describe a subclass of solutions for an abelian non-diagonal hYB deformation.  They have four free continuous complex variables (when solving for $\alpha_-$ they are $\epsilon_i$, $i=1, \ldots, 4$) and two free integer variables $m_1$ and $m_2$. These parameters should be subject to the extra constraints that the four charges $Q^L_{0,3},Q^R_{0,3}$ are real. To obtain the full resolvent we must substitute the solutions for $q_A^\pm, A, B, \alpha_\pm, $ and $n$ in \eqref{eq:ndaG1} and \eqref{eq:ndaG2}.

The undeformed limit is obtained by taking $\eta \rightarrow 0$ and, as we pointed out above, it would give us an additional constraint equation between parameters of the solution. Comparing this limit to solutions of the undeformed case may be subtle because here we took one of the branch points at $z=0$, by fixing $C=0$ from the beginning. The extra condition $C=0$ would therefore survive in the $\eta\to 0$ limit.

\subsection{Almost abelian}
In this section we explicitly construct the twist for the so-called almost abelian hYB deformations, following the terminology of~\cite{vanTongeren:2016eeb}. 
We will focus on rank-4 almost abelian deformations which means that we have solutions $R$ of the CYBE of the form 
\begin{equation}
R = R^{(1)} +R^{(2)} , 
\end{equation}
where each $R^{(i)} : \mathfrak{f}^\star_{(i)} \rightarrow \mathfrak{f}_{(i)}$ is a rank-2 abelian $R$-matrix. The deformation is \textit{almost abelian} when $\mathfrak{f} = \text{Im} (R)$ is \textit{non-abelian}, but  as a vector space we can write $\mathfrak{f} = \mathfrak{f}_{(1)} \oplus \mathfrak{f}_{(2)}$ with $\alg f_{(i)}$ abelian. In addition, $R^{(1)}$ must be compatible with the symmetries to construct $R^{(2)}$, meaning that if we write 
\begin{equation}
r^{(i)} = T^{(i)}_1 \wedge T^{(i)}_2 , \qquad [T_1^{(i)}, T_2^{(i)}] = 0 , \qquad T_{j}^{(i)} \in \mathfrak{f}_{(i)} ,
\end{equation}
then the compatibility reads
\begin{equation} \label{eq:compR1}
\text{ad}_{T_j^{(2)}} R^{(1)} = R^{(1)} \text{ad}_{T_j^{(2)}} , \qquad j=1,2 \ .
\end{equation}
Notice that this condition implies that $[\mathfrak{f}_{(2)}, \mathfrak{f}_{(1)}] \subset \mathfrak{f}_{(1)}$. With this information and using the Baker-Campbell-Hausdorff (BCH) formula,\footnote{In fact, from the BCH formula and $[\mathfrak{f}_{(2)},\mathfrak{f}_{(1)}]\subset \mathfrak{f}_{(1)}$ it follows that $\exp(-x_2)\exp(x_1+x_2)\in F_{(1)}$, and also $\exp(x_1+x_2)\exp(-x_2)\in F_{(1)}$, where $x_i\in \mathfrak{f}_{(i)}$.} one concludes that any $f\in F$ can be written as $f=f_1f_2$ or alternatively $f = f'_2f'_1$ with $f_i,f'_i\in F_{(i)}$, and $\mathfrak{f}_{(i)}=\text{Lie}(F_{(i)})$.\footnote{An example of an almost-abelian $R$-matrix is 
\begin{equation}
r = p_1\wedge p_2 + J_{12}\wedge p_3,
\end{equation}
where $p_i$ are generators of translations and $J_{12}$ of a rotation in  the algebra of the 3-dimensional Euclidean group, where $R$ satisfies the CYBE.  {One may actually  embed it in the conformal algebra to have a non-degenerate bilinear form}.
We can write this as the sum $r = r^{(1)} + r^{(2)}$ where $r^{(1)}=p_1\wedge p_2$, $r^{(2)}= J_{12}\wedge p_3$ and one may check that $R^{(1)}$ is compatible with the generators $ J_{12}, p_3$ in the sense defined above. For other examples see \cite{Borsato:2016ose,vanTongeren:2016eeb}. \label{f:AAexample}}

An interesting fact about almost abelian YB deformations is that they can be understood as the application of a non-commuting sequence of abelian deformations \cite{Borsato:2016ose,vanTongeren:2016eeb}. In the case that we are considering, they are equivalent to first applying the abelian hYB deformation with $R^{(1)}$, and later another abelian hYB deformation with $R^{(2)}$. The order of such sequence is important: if we  started by doing the deformation with $R^{(2)}$ first, in general we would not be able to apply the deformation with $R^{(1)}$, because the corresponding isometries may have already been broken. The compatibility~\eqref{eq:compR1} ensures that the sequence with first $R^{(1)}$ and then $R^{(2)}$ is always possible. We will now show that this mechanism is reflected in  the twist, which can be written as the product of the twists  for the abelian R-matrices $R^{(i)}$.

We start from the usual on-shell identification between the group element $g$ of hYB and a new group element $\tilde{\tilde g}$ of a twisted PCM
\be\label{eq:idgtt}
\tilde{\tilde g}^{-1} \partial_\pm \tilde{\tilde g} = \frac{1}{1\pm \eta R_g}g^{-1} \partial_\pm g .
\ee
We use a double tilde to indicate that we will interpret the deformation with $R$ as the composition of two deformations, and therefore as a double twist.
First we need the following identity
\be
\frac{1}{1\pm \eta R_g} = \frac{1}{1\pm \eta \left(1\pm \eta R^{(2)}_g\right)^{-1} R^{(1)}_g}\times \frac{1}{1\pm \eta R^{(2)}_g} \ ,
\ee
which only uses standard identities of invertible matrices, and does not assume any other property (e.g.~CYBE) for $R^{(i)}$. Using this identity,~\eqref{eq:idgtt} becomes
\be\label{eq:idgtt2}
\tilde{\tilde g}^{-1} \partial_\pm \tilde{\tilde g} = \frac{1}{1\pm \eta \left(1\pm \eta R^{(2)}_g\right)^{-1} R^{(1)}_g}
\tilde g^{-1} \partial_\pm \tilde g ,
\ee
where we introduced another group element $\tilde g$ that appears in the on-shell identification for the $R^{(2)}$ deformation as
\be
{\tilde g}^{-1} \partial_\pm {\tilde g} = \frac{1}{1\pm \eta R^{(2)}_g}g^{-1} \partial_\pm g .
\ee
This last equation is in fact what we would write if we were looking at the $R^{(2)}$ deformation only. Let us focus on this for a moment. We know that we can introduce a twist field $\mathcal F_{(2)}\in F_{(2)}$ such that
\be \label{eq:aabc1}
g = \mathcal F_{(2)} \tilde g,
\ee
and that we can write
\begin{equation} \label{eq:dF2AA}
\partial_\pm {\cal F}_{(2)} {\cal F}_{(2)}^{-1} = \pm \eta R^{(2)} \text{Ad}_{{\cal F}_{(2)}} \partial_\pm \tilde{g} \tilde{g}^{-1} \ ,
\end{equation}
similarly to \eqref{eq:dF2}.
As before, the periodicity of $g$ implies that $\tilde g$ has twisted boundary conditions
\be
\tilde g(2\pi) = W_{(2)} \tilde g (0) , \qquad W_{(2)} = {\cal F}_{(2)}^{-1} (2\pi) {\cal F}_{(2)} (0) \ .
\ee
Because $R^{(2)}$ is rank-2 abelian we know from  section~\ref{sec:ab} that the twist is simply
\be
W_{(2)}= \exp\left({-\eta R^{(2)}{\cal Q}^{(2)}}\right),\qquad
{{\cal Q}^{(2)}} = \int_0^{2\pi}d\sigma\ P_{(2)}^T(\partial_\tau {\tilde g}{\tilde g}^{-1} ).
\ee
Let us now go back to~\eqref{eq:idgtt2}. When using  $g = \mathcal F_{(2)} \tilde g$ we see that we can effectively just replace $g$ by $\tilde g$, {because the $\mathcal F_{(2)}$ dependence disappears}.\footnote{{In fact, $\AD_{\mathcal F_{(2)}}^{-1}R^{(2)}=R^{(2)}$ because $\alg f_{(2)}$ is abelian, and by antisymmetry of the $R$-matrix it also follows $R^{(2)}\AD_{\mathcal F_{(2)}}=R^{(2)}$.  Because of the assumption of compatibility in~\eqref{eq:compR1}, we also have $\text{Ad}_{\mathcal F_{(2)}}^{-1} R^{(1)} \text{Ad}_{{\mathcal F}_{(2)}} = R^{(1)}$.}}
Now we further identify 
\be \label{eq:aabc2}
\tilde g = \mathcal F_{(1)}\tilde{\tilde g},
\ee
via a new twist field $\mathcal F_{(1)}\in F_{(1)}$.\footnote{The fact  that $\mathcal F_{(1)}\in F_{(1)}$ follows from~\eqref{eq:1pR2f1}. In fact, from its right-hand-side we conclude that the expression is in the algebra $\alg f_{(1)}$. Now, say that $\partial_\pm \mathcal F_{(1)} \mathcal F_{(1)}^{-1} \in \alg h \subset \alg g$, the fact  that $\alg h=\alg f_{(1)}$ follows from noticing that  in the left hand side  we have two linear operators
$O_\pm = 1 \pm \eta R^{(2)}$
whose image is $\alg f_{(1)}$ when restricted to $\alg h$. The same must be true also for their semisum $(O_+ + O_-)/2 = 1$ which is the identity operator, and that is consistent only if we have $\alg h = \alg f_{(1)}$. This proves that $\mathcal F_{(1)}\in F_{(1)}$ (up to  multiplications by constant elements of $G$ from the right, that  can always be removed by redefining $\tilde{\tilde g}$.)} Using also \eqref{eq:dF2AA}, after some rewriting  we find that \eqref{eq:idgtt2} becomes
\be\label{eq:1pR2f1}
\left(1\pm \eta R^{(2)} 
\right) \partial_\pm \mathcal F_{(1)}\mathcal F_{(1)}^{-1} = \pm \eta R^{(1)} \AD_{\mathcal F_{(1)}}\partial_\pm \tilde{\tilde g}\tilde{\tilde g}^{-1}.
\ee
This looks very similar to~\eqref{eq:dF2}, with the exception of the additional term with $R^{(2)}$ on the left-hand-side.  We will now continue with the working assumption that $R^{(2)} \alg f_{(1)}=0$ which, in fact, holds for many of the relevant almost abelian examples (including the one mentioned in footnote \ref{f:AAexample}). Then the above equation becomes simply
\be
\mathcal F_{(1)}^{-1} \partial_\pm \mathcal F_{(1)} = \pm \eta R^{(1)}_{\mathcal F_{(1)}}\partial_\pm \tilde{\tilde g}\tilde{\tilde g}^{-1},
\ee
and we solve it in the same way as~\eqref{eq:dF2}. Since $R^{(1)}$ is also rank-2 abelian, the related twist is again simply
\be
W_{(1)}= \exp\left({-\eta R^{(1)} {\cal Q}^{(1)}}\right)
,\qquad
{{\cal Q}^{(1)}} = \int_0^{2\pi}d\sigma\ P_{(1)}^T(\partial_\tau \tilde {\tilde g}\tilde {\tilde g}^{-1} ).
\ee
Let us  now  look at the boundary conditions for $\tilde{\tilde g}$. Using \eqref{eq:aabc1}, \eqref{eq:aabc2} and $g(2\pi) = g(0)$, we find
\be\begin{aligned}
\tilde{\tilde g}(2\pi) 
&= W\ \tilde{\tilde g}(0),\qquad\qquad
W\equiv \mathcal F_{(1)}^{-1}(2\pi)\cdot W_{(2)}\cdot \mathcal F_{(1)}(0).
\end{aligned}
\ee
This twist has not immediately the form of the product of the two more elementary twists $W_{(i)}$. However, we can do the field redefinition $\tilde{\tilde g}(\sigma)\to \tilde{\tilde g}'(\sigma) =\mathcal F_{(1)}(2\pi)\ \tilde{\tilde g}(\sigma)$ which is just a reparameterisation of coordinates in the group manifold, and we obtain
\be
\tilde{\tilde g}'(2\pi) 
= W'\ \tilde{\tilde g}'(0),\qquad\qquad
W'=W_{(2)}\cdot W_{(1)},
\ee
where we also used $\mathcal F_{(1)}(0)\mathcal F_{(1)}^{-1}(2\pi)=\mathcal F_{(1)}^{-1}(2\pi)\mathcal F_{(1)}(0)$ because $\alg f_{(1)}$ is abelian.\footnote{Because the algebra $\alg f$  is non-abelian, $W'$ {cannot} be written as $W_{(1)}\cdot W_{(2)}$.
On the other hand, we could do a further field redefinition $\tilde{\tilde g}'(\sigma)\to \tilde{\tilde g}''(\sigma) =W_{(1)}\ \tilde{\tilde g}'(\sigma)$ and then the twisted boundary conditions would read $\tilde{\tilde g}''(2\pi) 
= W''\ \tilde{\tilde g}''(0)$ where
$W''=W_{(1)}\cdot W_{(2)}$. These possibilities are compatible because they correspond to different field parameterisations.} We conclude that, also at the level of the twisted model, almost abelian deformations have the nice feature of being understood as the application of a sequence of abelian twists. A different attempt to show this was made also in \cite{vanTongeren:2018vpb}.

{Notice that, if we insist on writing $W$ in factorised form, we are forced to write an expression in terms of $W_{(2)}$ which in general is local in terms of $\tilde g$, but may be non-local in terms of $\tilde{\tilde g}$. However, we already know from section~\ref{sec:twist} that the solution for $W$ also admits an expression that is perfectly local in $\tilde{\tilde g}$.}

Genuine almost abelian deformations will be compositions involving at least a non-diagonal abelian $R^{(1)}$ deformation,  since for a diagonal $R^{(1)}$ deformation the compatibility condition \eqref{eq:compR1} would imply that also the $R^{(2)}$ deformation is diagonal and thus the composed deformation would be abelian. This is in conflict with our initial assumptions.
Therefore, almost abelian deformations are characterised by non-diagonalisable twists that can be put at most in Jordan block form. The asymptotic behaviour of the quasimomenta will then be qualitatively similar to~\eqref{eq:p-as-non-diag}, with $\sqrt{z}$ terms. Therefore, the same comments made for the abelian non-diagonal twists hold, and we will not discuss this case further.

\subsection{Jordanian}
As a final example of the construction of the twist and the CSC, we now look at Jordanian deformations \cite{Kawaguchi:2014qwa,vanTongeren:2015soa,Hoare:2016hwh,Orlando:2016qqu} of a non-compact PCM model, and in particular of $\mathfrak{sl}(2,\mathbb{R}) \subset \mathfrak{g}$. Jordanian deformations illustrate  truly non-abelian hYB deformations, that cannot be understood as a sequence of abelian ones.  We take the choice and conventions of generators as in appendix \ref{app:CSC}, i.e. $T_a \in \mathfrak{sl}(2,\mathbb{R})$ with
\begin{equation}
[T_3 , T_\pm ] = \pm T_\pm , \qquad [T_+ , T_-] = 2 T_3 \ ,
\end{equation}
and for $T^a$ we have $T^3 = 2 T_3$ and $ T^\pm = T_\mp$. \\

\noindent {\bf The twist $W$} --- To work with the Jordanian deformation let us take 
\begin{equation}
r = T_+ \wedge T_3 ,
\end{equation}
which solves the CYBE, so that $\alg f$ is the Borel algebra generated by $T_+, T_3$. We will now show how to invert~\eqref{eq:red-y-X} exactly also in the Jordanian case. First decompose
\be
X = X_3 T^3+X_+T^+,\qquad\qquad
Y = Y_3T^3+Y_+T^+.
\ee
We have $RX=-X_+T_3+ X_3T_+$ and one can prove (for example by induction) that
\be
\text{ad}_{RX}^nX = X_+^{n+1} T^++ X_3 X_+^nT^3+f_nT^-,
\ee
where
\be
f_n=\left\{
\begin{array}{cl}
0, & n \text{ even}, \\
-2X_3^2 X_+^{n-1}, & n \text{ odd}.
\end{array}\right.
\ee
From~\eqref{eq:red-y-X} it then follows that 
\be
\eta Y = 
(1-e^{-X_+}) T^+ + X_3\ \frac{1-e^{-X_+}}{X_+} T^3,
\ee
which can easily be inverted to give
\be
X_+ =  - \log(1- \eta Y_+),\qquad \qquad
X_3 = - Y_3\frac{\log(1- \eta Y_+)}{Y_+}.
\ee
We now use the following identities coming from the BCH formula
\be
\exp(A)\exp(B) = \exp\left(A+\frac{s}{1-e^{-s}}B\right),\quad\text{and}\quad
\exp(A)\exp(B) \exp(-A)=\exp\left(e^sB\right),
\ee
which hold when for  $A,B \in \mathfrak{g}$ we have $[A,B]=sB$, with $s\neq 2\pi i n$. We can then rewrite the (inverse of the) twist field as
\be
\begin{aligned}
\mathcal F^{-1}  = \exp(-RX)
&=\exp(X_+ T_3)\exp\left(X_3\frac{e^{-X_+}-1}{X_+}T_+\right)\\
&=\exp\left( -\log(1-\eta Y_+)T_3\right)\exp\left(-\eta Y_3 T_+\right).
\end{aligned}
\ee
After some simple algebra, the twist is now found to be
\be
\begin{aligned}
W &= \mathcal F^{-1}(2\pi)\mathcal F(0)
=\exp\left(\log Q_A(T_3-Q_B T_+)\right) \ ,
\end{aligned}
\ee
where
\be
Q_A \equiv \frac{1-\eta\,   Y_+(0)}{1-\eta\,   Y_+(2\pi)},\qquad\qquad
Q_B \equiv \frac{Y_3(2\pi)-Y_3(0)}{Y_+(2\pi)-Y_+(0)},
\ee
are two conserved charges, because the twist is constant. In matrix form the twist reads
\begin{equation} \label{eq:twistJor}
W
=
\left(
\begin{array}{cc}
 \sqrt{Q_A} & \frac{Q_B-Q_A Q_B}{\sqrt{Q_A}} \\
 0 & \frac{1}{\sqrt{Q_A}} \\
\end{array}
\right).
\end{equation}
Let us point out that $W$ is always diagonalisable.
Note furthermore that the undeformed limit $\eta \rightarrow 0$ is simply obtained by sending $Q_A \rightarrow 1$.
The charge $Q_B$ can be understood as the ratio\footnote{Notice that these $\check Q^L_i$ have local expressions in terms of $\tilde g$, the field of the twisted PCM, which does not necessarily mean that they will be local in terms of $g$, the field of the hYB model. }
\be \label{eq:QBdef2}
Q_B = \frac{\check{Q}^L_3}{\check{Q}^L_+},\qquad \check{Q}^L_i =  \int_0^{2\pi}d\sigma\, \Tr\left(T_i\, \partial_\tau \tilde g\tilde g^{-1}\right).
\ee
In other words, even though the quantities $\check{Q}^L_3 , \check{Q}^L_+$ are conserved only in the $\eta\to 0$ limit and each of them is broken by the Jordanian deformation/twist, their ratio remains constant even in the presence of the deformation/twist. \\

\noindent {\bf Asymptotics of the quasimomenta} --- As before, the expansion of the quasimomenta at infinity does not change and thus will go as \eqref{eq:SL2QMInf}. However, at $z=0$ the twist contributes, and the quasimomenta will change compared to the  undeformed $\mathfrak{sl}(2,\mathbb{R}) $ case of appendix \ref{app:CSC}. 
Using the explicit expression of the Jordanian twist \eqref{eq:twistJor}
we find that  $\Omega'(z) = W^{-1} (1 - z \check{Q}^L ) + {\cal O}(z^2)$
at leading order around $z=0$ has two distinct eigenvalues, ${Q_A}^{-1/2}$ and ${Q_A}^{1/2}$, and that at the ${\cal O}(z)$ order,  they are proportional to
 $\check{Q}^L_3 - Q_B \check{Q}^L_+$. Hence, upon \eqref{eq:QBdef2}, the subleading order vanishes. Within the discussion of section \ref{sec:yb-int}, this means that the Jordanian deformation falls in a special class where one coordinate in the moduli space of Riemann surfaces of~\cite{Krichever:1996ut} is always zero, and it does not depend on conserved charges of the integrable model. To conclude, the quasimomenta at $z=0$ behave as\footnote{Recall that, for $\mathfrak{g}=\mathfrak{sl}(2, \mathbb{R})$ and our choice of algebra representation, the two eigenvalues of $\Omega'(z)$ are related as $\lambda_1 (z) = \lambda_2 (z)^{-1}$ and thus $p_1(z) = - p_2(z) \equiv p(z)$.}
\begin{equation}
p(z) \approx 2\pi  m + \frac{i}{2} \log Q_A +\mathcal O(z^2), \qquad z \approx 0 \ .
\end{equation}

Taking the undeformed limit at the level of the asymptotics of the quasimomenta is a subtle issue. In particular, the undeformed limit and the $z$-expansion of  the eigenvalues of $\Omega'(z)$ are two operations that do not commute. As mentioned, first expanding in $z$ does not lead to an ${\cal O}(z)$ contribution. If we first send $Q_A \rightarrow 1$, however, then we discover that the eigenvalues expand as $1 \pm  z\sqrt{(\check{Q}^L_3)^2 + \check{Q}^L_+ \check{Q}^L_- } + {\cal O}(z^2)$, as is expected for the undeformed case. This feature can be understood from the fact that, for general $Q_A$, the eigenvalues have a square root term of the form $\sqrt{Q^2_A -2 Q_A +1  + \# z^2 + \ldots}$. The $z$-term in the expansion then only appears when we take $Q_A \rightarrow 1$. Let us point out that even though we loose the ${\cal O}(z)$ contribution in the deformed/twisted case, some information on the conserved charges is, loosely speaking, restored through the appearance of the twist $W$ at leading order around $z=0$. \\

\noindent \textbf{One-cut resolvent} --- For the resolvent $G(z)$ capturing the class of solutions described by Riemann surfaces with single branch cuts of finite size we take an identical ansatz as given in \eqref{eq:SL2RG}. From demanding analyticity of $G(z)$ we then find the same solutions for $B$ and $C$ given in \eqref{eq:SL2RSolBC} and since also the asymptotics at infinity is identical we will have the same solution for $A$ as well, given in \eqref{eq:SL2RSolA}. Now, because the charge $Q_A$ appears only in the leading part of the resolvent around $z=0$, we will need to solve this equation for $Q_A$ rather than considering it as an equation which determines the charge ${\cal J}$, as we did for the undeformed case. We find
\begin{equation}\label{eq:solQA}
\log Q_A =  2 \pi i(2m - n) - 2 i \sqrt{C}  \frac{\kappa_- \epsilon_1 - \kappa_+ \epsilon_2}{\epsilon_1 \epsilon_2} \ ,
\end{equation}
where $C = C( \epsilon_i , \kappa_\pm, n)$. 
It is now the ${\cal O} (z)$ contribution of the resolvent around $z=0$ that gives a constraint equation. Indeed, using the solutions for $A,B$ and $C$ we find a quartic in $\epsilon_i$ but quadratic in $\kappa_\pm$ equation
\begin{equation}
\epsilon_1^4 \kappa_-^2 + 4 \epsilon_1^3 \epsilon_2 \kappa_+ \kappa_- + 4 \epsilon_1 \epsilon_2^3 \kappa_+ \kappa_- + \epsilon_2^4 \kappa_+^2 + \epsilon_1^2 \epsilon_2^2 \left( 3 (\kappa_+^2 + \kappa_-^2) - 4 n^2 \pi^2 \right) = 0 \ .
\end{equation}
Again this may be seen as an equation for the charge ${\cal J}$.
Finally, with the asymptotics at infinity being identical, we get the same solution for $Q_3^R$ as given in \eqref{eq:SL2RSolQR}. We conclude that we have two remaining free continuous complex variables, $\epsilon_1$ and $\epsilon_2$, which must be subject to extra conditions to ensure the reality of the charges\footnote{Note that the first imaginary term in~\eqref{eq:solQA} does not spoil the reality of $Q_A$.} $Q_A$ and $Q^R_3$, and two remaining free integers $n$ and $m$.

\section{Conclusions}\label{sec:concl}
In this paper we have considered the reformulation of periodic hYB models as undeformed yet twisted models. We provided a solution for the twist $W$ that is valid for the most generic hYB deformation, and that takes a local expression in terms of the fields of the twisted model. We then used this knowledge to discuss the CSC in some explicit cases.

In the case of diagonal abelian deformations it is known that, generalising what happens in the undeformed case, the finite-gap equations can be matched with a thermodynamic limit of the quantum asymptotic Bethe equations, when the number of Bethe roots goes to infinity and they condense into cuts~\cite{Frolov:2005ty,Beisert:2005if}. This knowledge may be in principle reversed to guess the Bethe equations from the finite-gap ones. We hope that our analysis of the CSC for the hYB models will give hints for a possible quantum integrability description of the hYB models that are not of diagonal abelian type. As  could be expected already before our analysis, and differently from the  diagonal abelian case, these deformations will not be implemented straightforwardly at the level of the asymptotic and thermodynamic Bethe equations. In order to understand the quantum integrability description of these models, it is likely that more progress on the worldsheet will be needed, for example in the understanding of the scattering problem despite the breaking of the BMN isometries. The construction of deformed/twisted spin-chains may also prove to be useful, see e.g.~\cite{Guica:2017mtd}.

When the twist is non-diagonalisable because it has the structure of Jordan blocks (as in the cases of non-diagonal abelian and almost abelian hYB deformations), the study of the eigenvalues of the monodromy matrix shows that they (and  the corresponding quasimomenta) must have a square-root branch point at $z=0$ (see also appendix~\ref{app:JordanForm}). This feature was also observed in both the gauge theory and $\sigma$-model picture of  the null dipole-deformed $AdS_5/CFT_4$ correspondence in \cite{Guica:2017mtd,Ouyang:2017yko}. In fact,  square-root cuts appear  in the asymptotics of the Q-functions in the dual field theory regime, which is highly quantum from the $\sigma$-model perspective.  This fact strengthens the suggestion that also in the Quantum Spectral Curve  modified asymptotics should be present. On the one hand, as we demonstrate, it is possible to construct at least some explicit solutions of the CSC for these deformed models with non-diagonalisable twists. On the other hand, because of the branch point at $z=0$, we are not able to claim that the data obtained by studying the quasimomenta around $z=0$ and $z=\infty$ is enough to reconstruct (at least in principle) the most generic class of finite-gap solutions. It would be interesting to have a better understanding on this issue, and the study of explicit solutions of the  equations of motion of the twisted  $\sigma$-model (that are supposed to be classified by the finite-gap solutions) may help.

It would also be interesting to further study  the case of Jordanian deformations. The twists in these cases are in fact diagonalisable, and they do not lead to the above complications. Moreover, the Jordanian deformations are viable candidates of deformations of AdS/CFT since they can be embedded in supergravity, given that unimodular~\cite{Borsato:2016ose} Jordanian $R$ matrices are possible in the case of superalgebras, as demontrated in~\cite{vanTongeren:2019dlq}. The Jordanian deformations that we consider here can then be viewed as truncations of the $\sigma$-models where fermions are set to zero. Given that the undeformed limit is straightforward at the level of the monodromy matrix, but  becomes more subtle at the level of its eigenvalues, it would be interesting to understand this further, with the generalisation to the quantum integrability description in mind.

{Finally, it would be nice to generalise our results to more generic deformations that involve the left and right copy of the symmetry group $G$ at the same time, see e.g.~\cite{Klimcik:2014bta}.}

We hope to come back to some of these issues in the future.

\section*{Acknowledgements}
We thank Tim Hollowood and Stijn van Tongeren for discussions, Stijn van Tongeren for comments on the manuscript, and Filip Jurukovi\'c for collaboration at the beginning of this project. RB and SD are supported by the fellowship of ``la Caixa Foundation'' (ID 100010434) with code LCF/BQ/PI19/11690019.
RB, SD and JLM are also supported by AEI-Spain (under project PID2020-114157GB-I00 and Unidad de Excelencia Mar\'\i a de Maetzu MDM-2016-0692),  Xunta de Galicia (Centro singular de investigaci\'on de Galicia accreditation 2019-2022, and project ED431C-2021/14), and by the European Union FEDER.

\appendix

\section{Classical spectral curves  for $U(2)$ and ${SL}(2,\mathbb R)$ PCM}\label{app:CSC}
In this appendix, we will describe the construction of the CSC for the undeformed $U(2)$ and $SL(2,\mathbb R)$ PCM. We will  show how to reconstruct  the quasimomenta, or equally the resolvents, in the case of the so-called one-cut solutions by  following~\cite{Kazakov:2004qf,Kazakov:2004nh}. Our  presentation will include some additional material, however, because  in the compact case we want to treat the larger $U(2)$ group, and because for $SL(2,\mathbb{R})$ we consider a small generalisation in the ansatz for the resolvent. This discussion will set the stage for the results of section~\ref{sec:ex}. 

 The PCM on the Lie groups $(i)\ G=U(2) = SU(2) \times U(1)$ and $(ii)\ G=SL(2, \mathbb{R})$ have target-space $(i)$ $S^3 \times S^1$ and $(ii)$ $AdS_3$, respectively.
For the generators $T_a \in \mathfrak{g}$ we take
\begin{alignat}{3}
(i)\;\; \mathfrak{g} &= \mathfrak{u}(2) : \quad &&T_0 = {\frac{i}{2}}\mathbf{1}, \; T_\alpha ={\frac{i}{2}} \sigma_\alpha , \;\; \alpha=1,2,3, \quad  &&T^\alpha = -2 T_\alpha, \label{eq:GenU2}  \\
(ii)\;\; \mathfrak{g} &= \mathfrak{sl}(2,\mathbb{R}): \quad  && T_\pm  = \frac{1}{2} (\sigma_1 \pm i \sigma_2) , \; T_3 = \frac{1}{2}\sigma_3 , \quad &&T^{3} = 2 T_{3}, \; T^\pm = T_\pm  \label{eq:GenSL2}\ ,
\end{alignat}
so that for \textit{(i)} they satisfy $T_a^\dagger=-T_a$, while for \textit{(ii)} they are real $T_a^\ast=T_a$.
In both cases we expand algebra elements as $x=x^a T_a$, with the components  $x^a$ real, and the group element is obtained by taking $\exp(x)$. Let us finally recall that we define the quasimomenta $p_A(z)$ as $\lambda_A(z) = e^{ip_A(z)}$ and their residues as in~\eqref{eq:defm}.

In this paper we are not introducing a worldsheet metric for the PCM action. One may, however, interpret our set-up as coming from  fixing  the conformal gauge when a worldsheet metric is present. In that case, one should additionally impose  Virasoro constraints.
We then have to make important distinctions between the  cases of $U(2)$ and $SL(2,\mathbb R)$. In particular, since $U(2)$ ($S^3 \times S^1$) describes a compact target space, we will add a time direction, so that we have $S^3 \times S^1\times \mathbb{R}$ with $\mathbb{R}$ parametrised by  target time $X_0$.
 On the other hand, since $SL(2,\mathbb{R})$ ($AdS_3$) already contains a time direction, we will add only a compact $U(1)$ ($S^1$) direction, so that we have $AdS_3 \times S^1$ with $S^1$ parametrised by an angle $\phi$.
Furthermore, let us point out that in the deformed examples of section \ref{sec:ex} we are applying an abelian deformation to the (full) $U(2)$ space, mixing its two Cartan directions. For that reason we will not decouple $U(2)$ in its non-abelian and abelian part here.
Respectively the PCM actions now simply are
\begin{alignat}{3}
(i) \; \;&U(2)\times \mathbb{R}: \quad &&S_{\text{PCM}} = -T \int d^2\sigma \left( \Tr J_+ J_- + \partial_+ X_0 \partial_- X_0  \right) , \quad  &&J_\pm \in \mathfrak{u}(2)\\
 (ii) \; \;&SL(2,\mathbb{R}) \times U(1): \quad &&S_{\text{PCM}} = T \int d^2\sigma \left( \Tr J_+ J_- + \partial_+ \phi \partial_- \phi  \right) , \quad &&J_\pm \in \mathfrak{sl}(2,\mathbb{R}) \ ,
\end{alignat}
and they should be supplemented with the Virasoro constraints,
\begin{alignat}{2}
(i)\;\; &\Tr J_+^2 + (\partial_+ X_0)^2 =  \Tr J_-^2 + (\partial_- X_0)^2 =0  , \qquad &&J_\pm \in \mathfrak{u}(2)\\
(ii)\; \;&\Tr J_+^2 + (\partial_+ \phi)^2 =  \Tr J_-^2 + (\partial_- \phi)^2 =0,  \qquad &&J_\pm \in \mathfrak{sl}(2,\mathbb{R}) \ .
\end{alignat}
We choose the gauges
\begin{equation}
\begin{aligned}
(i)\;\; X_0 = \kappa \tau , \qquad (ii)\; \;\phi = {\cal J} \tau + w \sigma ,
\end{aligned}
\end{equation}
with $w\in \mathbb{Z}$ a winding number around S$^1$, and $\kappa$ and ${\cal J}$  constants related to conserved charges of the system, see e.g.~\cite{Kazakov:2004qf,Kazakov:2004nh}.
In both cases, let us denote the eigenvalues of $J_\pm$ with $\lambda_A^{J_\pm}$. The Virasoro constraints now read
\begin{alignat}{3}
(i)\; \;  &(\lambda_1^{J_\pm})^2 + (\lambda_2^{J_\pm})^2 = - \kappa^2 , \label{eq:u2vir}\\
(ii) \; \; &(\lambda_1^{J_\pm})^2 + (\lambda_2^{J_\pm})^2 = 2(\lambda^{J_\pm})^2 =-\left( {\cal J} \pm w \right)^2 \label{eq:sl2vir},
\end{alignat}
where for the latter case we used that $J_\pm \in \mathfrak{sl}(2,\mathbb{R})$ is traceless and thus $ \lambda^{J_\pm}\equiv\lambda_1^{J_\pm} = -\lambda_2^{J_\pm} $. {In both cases notice that  the residues \eqref{eq:ResiduesQM} of the quasimomenta are, in general, distinct.}

\noindent {\bf Asymptotics of the quasimomenta} --- At infinity the monodromy matrix goes like \eqref{eq:Omega-inf} where $Q^R$ is defined in \eqref{eq:qr0}. Using the global $G_R$-invariance of the PCM allows us to diagonalise $Q^R$ such that
\begin{equation}
\begin{aligned}
(i) \;\; Q'^R &=  Q'^R_0 T^0 + Q'^R_3 T^3  , \qquad
(ii) \;\; Q'^R = { Q'^R_3 }T^3 \ .
\end{aligned}
\end{equation}
From now on we will assume that we have implemented the diagonalisation and thus omit the primes. Therefore, for the quasimomenta  we have
\begin{alignat}{2} 
(i) \;\; &p_1(z) \approx -\frac{1}{z} \left( Q^R_0 + Q^R_3 \right) , \qquad p_2(z) \approx -\frac{1}{z} \left( Q^R_0 - Q^R_3 \right) , \qquad &&z \approx \infty , \label{eq:U2QMInf}\\
(ii)\;\; &p(z) = p_1(z) = - p_2(z) \approx -\frac{i}{z}  Q^R_3 , \qquad &&z \approx \infty \ . \label{eq:SL2QMInf}
\end{alignat} 
Around $z=0$ the reasoning is the same. The monodromy matrix behaves as \eqref{eq:Omega-0} where in the undeformed case $W=1$ and $Q^L = \check{Q}^L$, defined in \eqref{eq:ql0}, is conserved. The global $G_L$-invariance of the PCM allows us to diagonalise $Q^L$ and thus we can conclude that the quasimomenta behave as
\begin{alignat}{2} 
(i) \;\; &p_1(z) \approx 2\pi m_1 + z \left( Q^L_0 + Q^L_3 \right) , \qquad p_2(z) \approx 2\pi m_2 + z \left( Q^L_0 - Q^L_3 \right) , \qquad &&z \approx 0 , \label{eq:U2QM0} \\
(ii)\;\; &p(z) = p_1(z) = - p_2(z) \approx 2\pi  m + iz Q^L_3 , \qquad &&z \approx 0 \ , \label{eq:SL2QM0}
\end{alignat}
with $m,m_1, m_2 \in \mathbb{Z}$.
According to the discussion of section \ref{sec:yb-int}, the above data---together with the residues \eqref{eq:ResiduesQM} around $z = \pm 1$ and the Riemann-Hilbert problems in \eqref{eq:ppn}---is enough to reconstruct the full Riemann surface {and} the quasimomenta/resolvents, once the location and number of the cuts have been specified.

\noindent {\bf One-cut resolvents} --- In the remaining part of this appendix we will illustrate the reconstruction of the resolvent defined in \eqref{eq:defResolvent} for the class of solutions defined by one single cut of finite size.
We will show in this case how this method can express the  charges   of the theory in terms of  variables {which could be related to variables of the sigma-model solutions}. 
  The Riemann-Hilbert problems \eqref{eq:ppn} in terms of the resolvents simply become
\begin{equation} \label{eq:resolventsheet}
\begin{aligned}
G_1 (z +i 0) - G_2 (z-i0) = 2\pi n -    \frac{q_1^+ - q_2^+}{1 - z}  +  \frac{q_1^- - q_2^-}{1 + z}, \qquad z \in \text{Cut} ,
\end{aligned}
\end{equation}
with $G_2(z) = -G_1(z)$ for the $SL(2,\mathbb{R}) $ case.  Let us now treat case \textit{(i)} and case \textit{(ii)} separately. \\

 \textbf{\textit{(i)}} $\mathbf{U(2)}$ --- Note that,  after diagonalisation,  the quasimomenta for a rank two PCM model will have the form $p_1(z) = f(z) + \sqrt{g(z)}$ and $p_2(z) = f(z) - \sqrt{g(z)}$, for some function $f(z)$ with single poles at $z = \pm 1$ and a regular part elsewhere, and a polynomial function $g(z)$ defining the finite size branch cuts. Therefore, for the one-cut resolvents let us take the ansatz
\begin{align} \label{eq:U2G1}
G_1(z) &= \pi n + \left(\frac{\alpha_+ - q_1^+}{1 - z}  +  \frac{\alpha_- + q_1^-}{1 + z} \right) -   \left(\frac{ \epsilon_1^{-1} q_1^+}{1 - z}  -  \frac{\epsilon_2^{-1} q_1^-}{1 + z} \right) \sqrt{Az^2+Bz+C} , \\
G_2(z) &= -\pi n + \left(\frac{\alpha_+ - q_2^+}{1 - z}  +  \frac{\alpha_- + q_2^-}{1 + z} \right) +  \left(\frac{ \epsilon_3^{-1} q_2^+}{1 - z}  -  \frac{\epsilon_4^{-1} q_2^-}{1 + z} \right) \sqrt{Az^2+Bz+C} , \label{eq:U2G2}
\end{align}
with the constraint
\begin{equation} \label{eq:u2constraint}
\frac{ \epsilon_1^{-1} q_1^+}{1 - z}  -  \frac{\epsilon_2^{-1} q_1^-}{1 + z}  =\frac{ \epsilon_3^{-1} q_2^+}{1 - z}  -  \frac{\epsilon_4^{-1} q_2^-}{1 + z},
\end{equation}
 for all $z$. Then \eqref{eq:resolventsheet} will be automatically satisfied.
Here, $\alpha^\pm$, $\epsilon_1, \epsilon_2,\epsilon_3,\epsilon_4$, $A, B$ and $C$ are all parameters. Some of them will be fixed by requiring the correct behaviour of the resolvents, the others {may be interpreted as  free parameters} for the sigma model solution. 

We can now solve \eqref{eq:u2constraint}  simply by $q_1^+ = \frac{\epsilon_1}{\epsilon_3} q_2^+$ and $q_1^- = \frac{\epsilon_2}{\epsilon_4} q_2^-$. {Note that if  $\epsilon_1 = -\epsilon_3$ and $\epsilon_2 = - \epsilon_4$ then $J_\pm$ is in $\mathfrak{su}(2)\subset\alg u(2)$ and we would not be describing $\alg u(2)$ in full generality.} There are several additional requirements that the resolvents must satisfy. First they should be analytical everywhere, and thus the contributions from $(1\pm z)^{-1}$ are required to cancel. This implies\footnote{Let us point out that the inclusion of the parameters $\alpha_\pm$ is essential here. When $\alpha_\pm = 0$ we see  that, in order to have non-trivial quantities $q_A^\pm$, we must take at the same time $\epsilon_1 = -\epsilon_3$ and $\epsilon_2 = -\epsilon_4$. As mentioned before, this is also not generic enough since it would describe only the $SU(2)$ subsector of our problem.}
\begin{equation} \label{eq:U2qpm}
q_1^+ =   \frac{2 \epsilon_1}{\epsilon_1 + \epsilon_3} \alpha_+ , \qquad q_1^- =  - \frac{2 \epsilon_2}{\epsilon_2 + \epsilon_4} \alpha_-, \qquad q_2^+ =  \frac{2 \epsilon_3}{\epsilon_1 + \epsilon_3} \alpha_+ , \qquad q_2^- = -  \frac{2 \epsilon_4}{\epsilon_2 + \epsilon_4} \alpha_- , 
\end{equation}
and
\begin{equation} \label{eq:U2SolBC}
B = \frac{1}{8} \left( \epsilon_1 + \epsilon_2 - \epsilon_3 - \epsilon_4\right) \left( \epsilon_1 - \epsilon_2 - \epsilon_3 + \epsilon_4 \right) ,\qquad C= \frac{1}{8} \left( (\epsilon_1 - \epsilon_3)^2 + (\epsilon_2 - \epsilon_4)^2 \right) - A .
\end{equation}
Secondly, let us solve the two equations {obtained from matching the ${\cal O}((1/z)^{0})$ contribution of \eqref{eq:U2QMInf} with the ansatze of the resolvents}. These in fact become identical  upon the solutions for $q_A^\pm$, and give
\begin{equation} \label{eq:U2SolA}
\sqrt{A} =  \frac{(\epsilon_1 + \epsilon_3) ( \epsilon_2 + \epsilon_4 )}{2(\epsilon_1 + \epsilon_3) \alpha_- - 2(\epsilon_2 + \epsilon_4) \alpha_+ } n \pi \ .
\end{equation}
Similarly, we can solve  the leading order ${\cal O}(z^0)$  of \eqref{eq:U2QM0} in terms of, for instance, $\alpha_+$ and $\sqrt{C}$. We find
\begin{align} \label{eq:U2Solap}
\alpha_+ &=  (m_1 + m_2) \pi - \alpha_- , \\
\sqrt{C} &= \frac{(\epsilon_1 + \epsilon_3) (\epsilon_2 + \epsilon_4) (n-m_1+m_2) \pi}{2 (\epsilon_1 + \epsilon_3) \alpha_- + 2 (\epsilon_2 + \epsilon_4) (m_1 \pi + m_2 \pi - \alpha_-) }   ,  \label{eq:U2SolsC}
\end{align}
and comparing the result for $\sqrt{C}$
 with the solutions \eqref{eq:U2SolA} and \eqref{eq:U2SolBC} leads to the following quartic constraint equation
\begin{equation} \label{eq:U2CQ4}
\begin{aligned}
0 = &\left( (\epsilon_1 - \epsilon_3)^2 + (\epsilon_2 - \epsilon_4)^2 \right) \left( (\epsilon_1 + \epsilon_3)^2 \alpha_-^2 - (\epsilon_2 + \epsilon_4)^2 \alpha_+^2 \right)^2 \\
&- 2 (\epsilon_1 + \epsilon_3)^2 (\epsilon_2 + \epsilon_4)^2 n^2 \pi^2 \left( (\epsilon_1 + \epsilon_3) \alpha_- + (\epsilon_2 + \epsilon_4) \alpha_+ \right)^2 \\
&- 2 (\epsilon_1 + \epsilon_3)^2 (\epsilon_2 + \epsilon_4)^2 (n-m_1 + m_2)^2 \pi^2 \left( (\epsilon_1 + \epsilon_3) \alpha_- - (\epsilon_2 + \epsilon_4) \alpha_+ \right)^2  \ .
\end{aligned}
\end{equation}
Thirdly, we can solve the equations from \eqref{eq:U2QMInf} and  \eqref{eq:U2QM0}  linear in $z$ and $1/z$ respectively for  $Q^R$ and $Q^L$  as
\begin{equation} \label{eq:U2SolQR}
\begin{aligned}
Q_0^R &= Q_0^L = \alpha_+ - \alpha_- , \\
Q_3^R &= -\frac{2 A  \left( (\epsilon_2 + \epsilon_4) \alpha_+ + (\epsilon_1 + \epsilon_3 )\alpha_- \right) + B  \left( (\epsilon_2 + \epsilon_4) \alpha_+ - (\epsilon_1 + \epsilon_3 )\alpha_- \right)  }{(\epsilon_1 + \epsilon_3) (\epsilon_2 + \epsilon_4) \sqrt{A}}  , \\
Q_3^L &= -\frac{2 C \left( (\epsilon_2 + \epsilon_4) \alpha_+ - (\epsilon_1 + \epsilon_3 )\alpha_- \right) + B \left( (\epsilon_2 + \epsilon_4) \alpha_+ + (\epsilon_1 + \epsilon_3 )\alpha_- \right)  }{(\epsilon_1 + \epsilon_3) (\epsilon_2 + \epsilon_4) \sqrt{C}} , 
\end{aligned}
\end{equation}
in which the explicit expressions for $A,B,C$ and $\alpha_+$ (in terms of $\epsilon_i, m_1,m_2,n,\alpha_-$) must be substituted. In addition, they should be subjected to the quartic constraint \eqref{eq:U2CQ4} as well as the quadratic  Virasoro constraints, $(\lambda_1^{J_+})^2 + (\lambda_2^{J_+})^2 =(\lambda_1^{J_-})^2 + (\lambda_2^{J_-})^2$ implying a relation between $q_A^\pm (\alpha_\pm , \epsilon_i)$.
{Thus,  upon the constraint equations, we  have effectively three remaining free continuous variables within this class of sigma model solutions, for instance $\epsilon_i$ for $i=1,2,3$. One may fix these free variables in order to get more simpler expressions, see e.g. \cite{Kazakov:2004qf}. In addition, we have three  free integers, $n,m_1$ and $m_2$.}
{Notice how the constraints lead to the identificaiton of the left and right $U(1)$ charge ($Q_0^R = Q_0^L$), which is consistent with the fact that the actions of left and right $U(1)$ transformations must necessarily be the same.}\\

 \textbf{\textit{(ii)}} $\mathbf{SL(2,\mathbb{R})} $ --- In this case, we have the simplification $G_1(z) = - G_2(z) \equiv G(z)$ and $q_1^\pm = - q_2^\pm \equiv \kappa^\pm$.  We can solve for $\kappa^\pm$ immediately by using the Virasoro constraints which give $(\kappa^\pm)^2 = \frac{\pi^2}{2} ({\cal J}\pm w)^2$. This is in contrast to the previous case. For a one-cut resolvent a sufficient ansatz now is,
\begin{equation} \label{eq:SL2RG}
G(z)= \pi n - \left(\frac{\kappa^+}{1 - z}  -  \frac{ \kappa^-}{1 + z} \right) -   \left(\frac{ \epsilon_1^{-1} \kappa^+}{1 - z}  -  \frac{\epsilon_2^{-1} \kappa^-}{1 + z} \right) \sqrt{Az^2+Bz+C} .
\end{equation}
Following the same steps as before, we first require that the contributions from $(1\pm z)^{-1}$ cancel by setting
\begin{equation} \label{eq:SL2RSolBC}
B = \frac{1}{2} \left( \epsilon_1^2 - \epsilon_2^2 \right) , \qquad C = \frac{1}{2} \left( \epsilon_1^2 + \epsilon_2^2 \right) - A \ .
\end{equation}
Solving the constant part ${\cal O}((1/z)^0)$ of \eqref{eq:SL2QMInf} gives,
\begin{equation} \label{eq:SL2RSolA}
\sqrt{A} = - \frac{\epsilon_1 \epsilon_2 }{\epsilon_1 \kappa^- + \epsilon_2 \kappa^+} n \pi \ .
\end{equation}
From comparing the leading part of \eqref{eq:SL2QM0} with the solutions for $A$ and $C$ we again get a quartic constraint equation,
\begin{equation} \label{eq:SL2RC4}
\left(\epsilon_1^2 + \epsilon_2^2 \right) \left(\epsilon_1^2 (\kappa^-){}^2 - \epsilon_2^2 (\kappa^+){}^2 \right)^2 - 2 \epsilon_1^2 \epsilon_2^2 \left( n^2 \left( \epsilon_1 \kappa^- - \epsilon_2 \kappa^+ \right)^2 + (n-2  m)^2 \left( \epsilon_1 \kappa^- + \epsilon_2 \kappa^+ \right)^2  \right) \pi^2= 0
\end{equation}
which we can think of as an equation for the charge ${\cal J}$. Finally we can solve the linear parts in $1/z$ and $z$  of \eqref{eq:SL2QMInf} and \eqref{eq:SL2QM0} respectively for $Q^R_3$ and $Q^L_3$ as
\begin{align} \label{eq:SL2RSolQR}
Q^R_3 &= i  \frac{(B+2A) \epsilon_2 \kappa^+ + (B-2A) \epsilon_1 \kappa^-}{2 \sqrt{A} \epsilon_1 \epsilon_2} , \\
Q^L_3 &= i \frac{(B+2 C) \epsilon_2 \kappa^+ - (B-2 C) \epsilon_1 \kappa^-}{2 \sqrt{C} \epsilon_1 \epsilon_2} . \label{eq:SL2RSolQL}
\end{align}
Let us remark that here we  have effectively two remaining free continuous complex variables within this class of sigma model solutions, namely $\epsilon_1$ and $\epsilon_2$, and they may be subject to conditions such that the conserved charges obtained are real. One may further restrict them to be related in certain ways  to get simpler expressions, see e.g.~\cite{Kazakov:2004nh}. In addition we have two free integers, $n$ and $m$.

\section{Non-diagonalizable monodromy matrices and branch points}\label{app:JordanForm}

In this appendix we review some aspects of non-diagonalizable matrices and their relation with the branch points of their eigenvalues.

It is customary to assume that the monodromy matrix can be diagonalized to argue that equation~(2.30),
\be
\label{eq:Laxevol}
\partial_\tau\Omega(z,\tau) =-\mathcal L_\tau(z,\tau,2\pi)\Omega(z,\tau)+\Omega(z,\tau)\mathcal L_\tau(z,\tau,0),
\ee
together with the boundary conditions $\mathcal L_\tau(z,\tau,0)=\mathcal L_\tau(z,\tau,2\pi)$, ensures that its eigenvalues are conserved quantities. 
However, the solutions of the CYBE often lead to monodromy matrices that cannot be diagonalized when evaluated at special values of $z$. It is worth pointing out that the assumption of diagonalizability is not necessary to show that the  eigenvalues of the monodromy matrix are conserved quantities.
A more general statement can be established using that any complex matrix can be conjugated to its Jordan canonical form. Explicitly, this means that there is a non-singular matrix $U(\tau,z)$ such that
\be
U(z,\tau)\, \Omega(z,\tau)\, U^{-1}(z,\tau)= \text{diag}\big(J_{n_1}(\lambda_1),\ldots, J_{n_k}(\lambda_k)\big),
\ee
where
\be
J_{n}(\lambda)=\left(\begin{array}{cccc}\lambda & 1 &  &  \\ &  \ddots & \ddots&  \\   &  & \lambda & 1 \\   &  &  & \lambda\end{array}\right)
\ee
is an $n\times n$ Jordan block, and $\{\lambda_i\equiv \lambda_i(z,\tau),\; i=1,\ldots,k\}$ are the eigenvalues of $\Omega(z,\tau)$. The Jordan canonical form is unique up to a permutation of the Jordan blocks. Moreover, a matrix is diagonalizable if all the Jordan blocks are~$1\times1$.

Since~\eqref{eq:Laxevol} is a first order differential equation, and using that $\mathcal L_\tau(z,\tau,0)=\mathcal L_\tau(z,\tau,2\pi)$, its solution can be written as
\be
\label{eq:solOmega}
\Omega(z,\tau) = \Psi(z,\tau)\,  \Omega(z,0)\, \Psi^{-1}(z,\tau),
\ee
where $\Psi(z,\tau)$ is the unique solution of
\be
\partial_\tau \Psi(z,\tau) = -\mathcal L_\tau(z,\tau,2\pi)\, \Psi(z,\tau),\qquad  \Psi(z,0)=1\,.
\ee
Then, plugging the Jordan canonical form of $\Omega(z,0)$ into~\eqref{eq:solOmega}, one concludes that $\lambda_i(z,\tau)=\lambda_i(z,0)$, which confirms that the eigenvalues are conserved quantities.

It is also important to recall the relationship between the branch points of the eigenvalues in the complex $z$-plane and the diagonalizability of $\Omega(z)$. Following~\cite{Babelon:2003qtg} (Remark 2 in sec.~5.2), it is useful to understand this on the simple example of a $2\times2$ matrix like
\be
\Omega(z)=\left(\begin{array}{cc}a(z) & b(z) \\c(z)& d(z)\end{array}\right)\,.
\ee
It has eigenvalues
\be
\lambda_\pm(z)= \frac{a(z)+d(z)}2 \pm \frac{\sqrt{\Delta(z)}}2\,,\qquad \Delta(z)=(a(z)-d(z))^2 +4b(z) c(z)\,,
\ee
and (normalized) eigenvectors
\be
\Psi_\pm(z)=\left(\begin{array}{c}1 \\ \psi_\pm(z)\end{array}\right)\,,\qquad
\psi_\pm(z)=\frac{d(z)-a(z)}{2b(z)}\pm \frac{\sqrt{\Delta(z)}}{2b(z)}\,.
\ee
We will be interested in the situation where $z_0$ is a root of $\Delta(z)$, so that $\Delta(z_0)=0$ and at $z=z_0$ the two eigenvalues are equal,
\be
\lambda_+(z_0)=\lambda_-(z_0)= \frac{a(z_0)+d(z_0)}2\equiv \lambda_0\,.
\ee
Recall that a matrix can be diagonalized if there exists a basis of eigenvectors. Therefore, we can distinguish two cases:

\begin{itemize}

\item[a)]
$b(z_0)\not=0$ (or $c(z_0)\not=0$). In this case the two eigenvectors at $z=z_0$ are equal, 
\be
\psi_+(z_0)=\psi_-(z_0)=\frac{d(z_0)-a(z_0)}{2b(z_0)},
\ee
and one cannot construct a basis of eigenvectors. Therefore, $\Omega(z_0)$ is not diagonalizable and, instead, one can find an invertible matrix $U(z)$ such that it can be conjugated to its Jordan canonical form
\be
U^{-1}(z_0) \Omega(z_0) U(z_0)= \left(\begin{array}{cc}\lambda_0 & 1 \\0 & \lambda_0\end{array}\right)\,.
\ee
If $c(z_0)\not=0$, we reach the same conclusion starting with the equivalent set of eigenvectors\footnote{$\widetilde\Psi_\pm(z)= \psi^{-1}_\pm(z)\Psi_\pm(z)$ are the same eigenvectors normalized in a different way.}
\be
\widetilde\Psi_\pm(z)=\left(\begin{array}{c}\psi^{-1}_\pm(z) \\ 1\end{array}\right)\,,\qquad
\psi^{-1}_\pm(z)=-\frac{d(z)-a(z)}{2c(z)}\pm \frac{\sqrt{\Delta(z)}}{2c(z)}\,.
\ee

Let us consider the case $b(z_0)\not=0$. Assuming that $a(z)-d(z)$ and $c(z)$ vanish to first order in $z-z_0$, namely  $a(z)-d(z),\, c(z)\sim (z-z_0)$, which is equivalent to
\be
\Delta(z)\sim \delta (z-z_0)+\cdots
\ee
where $\delta$ is a constant, then
\be
\lambda_\pm (z)\sim \lambda_0 \pm \frac{1}2\sqrt{\delta}\; \sqrt{z-z_0}+\cdots\,,
\ee
exhibit a square-root branch point at $z=z_0$.\footnote{Of course, it could happen that $\Delta(z)\sim \delta (z-z_0)^n+\cdots$, which would lead to a different type of singularity even though the matrix is still not diagonalizable.} This particular assumption agrees with the structure of the monodromy matrix of the non-diagonal abelian example  around $z=0$, cf.~section \ref{ss:abnondiag}, according to equations~\eqref{eq:Omega-0} and~\eqref{eq:Wnda}.

\item[b)]
If both $b(z_0)=c(z_0)=0$ vanish, then $a(z_0)-d(z_0)=0$. If  we assume that
all of them vanish to first order in $z-z_0$, 
\begin{equation}\begin{split}
&b(z)=\beta(z-z_0)+\cdots, \qquad c(z)=\gamma(z-z_0)+\cdots, \quad\\[5pt]
&a(z_0)-d(z_0)=\alpha(z-z_0)+\cdots,
\end{split}\end{equation}
then the two eigenvectors tend to different limits at $z=z_0$
\be
\psi_\pm(z)= -\frac{\alpha}{2\beta} \pm \frac{\sqrt{\alpha^2+4\beta\gamma}}{2\beta}+{\cal{O}}(z-z_0)+\cdots\,,
\ee
if $\alpha^2+4\beta\gamma\neq 0$. In that case one can construct a basis of eigenvectors. Therefore, the matrix is diagonalizable, and one can find an invertible matrix $U(z)$ such that
\be
U^{-1}(z_0) \Omega(z_0) U(z_0)= \left(\begin{array}{cc}\lambda_0 & 0 \\0 & \lambda_0\end{array}\right)\,.
\ee
In this case,
\be
\lambda_\pm(z)\sim  
\lambda_0+ \frac{a'(z_0)+ d'(z_0)}{2}\, (z-z_0) 
\pm \frac{\sqrt{\alpha^2+4\beta\gamma}}{2}\; (z-z_0)+\cdots
\ee
and, therefore, $z=z_0$ is not a branch point.

\end{itemize}

In general, if $\lambda_+(z_0)=\lambda_-(z_0)$, which requires that $\Delta(z_0)=0$, the matrix will be diagonalizable provided that $b(z_0)=c(z_0)=0$ and
\be
\label{eq:zeros}
\lim_{z\to z_0} \frac{\sqrt{\Delta(z)}}{b(z)}\quad \text{and}\quad  \lim_{z\to z_0} \frac{\sqrt{\Delta(z)}}{c(z)} \not=0
\ee
and finite. Then, assuming that all the entries in $\Omega(z)$ can be expanded in powers of $z-z_0$, one gets
\be
\sqrt{\Delta(z)}, b(z), c(z) \sim (z-z_0)^n
\ee
for some positive integer $n$ and, thus, $z=z_0$ is not a branch point. 

At least in this case of $2\times 2$ matrices, branch points correspond to values of $z$ where two eigenvalues become degenerate and the monodromy matrix is non-diagonalizable. In any case, notice that one cannot ensure that a value of $z$ where two eigenvalues become degenerate and the monodromy matrix is non-diagonalizable will always correspond to a branch point.


\bibliographystyle{nb}
\bibliography{biblio}{}

\end{document}